%% file: 0_Main.tex
\documentclass[lettersize,journal]{IEEEtran}
\usepackage{cite}
\usepackage{amsmath,amsfonts}

\usepackage{amsmath,amssymb,amsfonts}
\usepackage{algorithmic}
\usepackage{graphicx}
\usepackage{textcomp}
\usepackage{soul}

\usepackage{hyperref}
\usepackage{makecell}
\usepackage{multirow}
\usepackage{array}
\usepackage{caption}
\usepackage{subcaption}
\usepackage{placeins}
\usepackage{tabularx}
\usepackage{booktabs}
\usepackage{tablefootnote}
\usepackage{color}
\usepackage[table]{xcolor}
\usepackage{amsmath}
\usepackage{amsmath}
\usepackage{array}

\newcolumntype{P}[1]{>{\centering\arraybackslash}p{#1}}
\newcolumntype{M}[1]{>{\centering\arraybackslash}m{#1}}

\def\BibTeX{{\rm B\kern-.05em{\sc i\kern-.025em b}\kern-.08em
    T\kern-.1667em\lower.7ex\hbox{E}\kern-.125emX}}
\begin{document}

\title{KVC-onGoing: Keystroke Verification Challenge}

\author{\IEEEauthorblockN{Giuseppe Stragapede\IEEEauthorrefmark{1}\thanks{Email: \href{mailto:giuseppe.stragapede@uam.es}{giuseppe.stragapede@uam.es}}, 
Ruben Vera-Rodriguez\IEEEauthorrefmark{1},
Ruben Tolosana\IEEEauthorrefmark{1},
Aythami Morales\IEEEauthorrefmark{1},
Ivan DeAndres-Tame\IEEEauthorrefmark{1},\\
Naser Damer\IEEEauthorrefmark{2},
Julian Fierrez\IEEEauthorrefmark{1},
Javier Ortega-Garcia\IEEEauthorrefmark{1},
Alejandro Acien\IEEEauthorrefmark{1},
Nahuel Gonzalez\IEEEauthorrefmark{3},
Andrei Shadrikov\IEEEauthorrefmark{4},\\
Dmitrii Gordin\IEEEauthorrefmark{5},
Leon Schmitt\IEEEauthorrefmark{6},
Daniel Wimmer\IEEEauthorrefmark{6},
Christoph Großmann\IEEEauthorrefmark{6},
Joerdis Krieger\IEEEauthorrefmark{6}, 
Florian Heinz\IEEEauthorrefmark{6},\\
Ron Krestel\IEEEauthorrefmark{6},
Christoffer Mayer\IEEEauthorrefmark{6},
Simon Haberl\IEEEauthorrefmark{6},
Helena Gschrey\IEEEauthorrefmark{6},
Yosuke Yamagishi\IEEEauthorrefmark{7},\\
Sanjay Saha\IEEEauthorrefmark{8}, 
Sanka Rasnayaka\IEEEauthorrefmark{8},
Sandareka Wickramanayake\IEEEauthorrefmark{9}, 
Terence Sim\IEEEauthorrefmark{8},
Weronika Gutfeter\IEEEauthorrefmark{10},\\
Adam Baran\IEEEauthorrefmark{10},
Mateusz Krzyszto\'n\IEEEauthorrefmark{10} and
Przemysław Jask\'oła\IEEEauthorrefmark{10}
}


\IEEEauthorblockA{\IEEEauthorrefmark{1}Biometrics and Data Pattern Analytics (BiDA) Lab, Universidad Autonoma de Madrid, Spain}\\
\IEEEauthorblockA{\IEEEauthorrefmark{2}Fraunhofer Institute for Computer Graphics Research IGD, Darmstadt, Germany}\\
\IEEEauthorblockA{\IEEEauthorrefmark{3}Laboratorio de Sistemas de Información Avanzados (LSIA), Universidad de Buenos Aires, Argentina}\\
\IEEEauthorblockA{\IEEEauthorrefmark{4}Verigram LLC, Singapore}\\
\IEEEauthorblockA{\IEEEauthorrefmark{5}Citix, Almaty, Kazakhstan}\\
\IEEEauthorblockA{\IEEEauthorrefmark{6}Regensburg University Of Applied Sciences, Germany}\\
\IEEEauthorblockA{\IEEEauthorrefmark{7}Division of Radiology and Biomedical Engineering, The University of Tokyo, Japan}\\
\IEEEauthorblockA{\IEEEauthorrefmark{8}National University of Singapore, Singapore}\\
\IEEEauthorblockA{\IEEEauthorrefmark{9}University of Moratuwa, Sri Lanka}\\
\IEEEauthorblockA{\IEEEauthorrefmark{10}NASK – National Research Institute, Warsaw, Poland}\\
}




\maketitle





\vspace{-21pt}

\begin{abstract}
This article presents the Keystroke Verification Challenge - onGoing (KVC-onGoing)\footnote{\href{https://sites.google.com/view/bida-kvc/}{https://sites.google.com/view/bida-kvc/}}, on which researchers can easily benchmark their systems in a common platform using large-scale public databases, the Aalto University Keystroke databases, and a standard experimental protocol. The keystroke data consist of tweet-long sequences of variable transcript text from over 185,000 subjects, acquired through desktop and mobile keyboards simulating real-life conditions. The results on the evaluation set of KVC-onGoing have proved the high discriminative power of keystroke dynamics, reaching values as low as 3.33\% of Equal Error Rate (EER) and 11.96\% of False Non-Match Rate (FNMR) @1\% False Match Rate (FMR) in the desktop scenario, and 3.61\% of EER and 17.44\% of FNMR @1\% at FMR in the mobile scenario, significantly improving previous state-of-the-art results. Concerning demographic fairness, the analyzed scores reflect the subjects’ age and gender to various extents, not negligible in a few cases. The framework runs on CodaLab\footnote{\href{https://codalab.lisn.upsaclay.fr/competitions/14063}{https://codalab.lisn.upsaclay.fr/competitions/14063}}.
\end{abstract}



\section{Introduction}
\input{1_Introduction}

\section{Related Work}
\label{sec:related_work}

\input{2a_Related_Work}

\section{Resources Provided}
\label{sec:resources}
\input{2_Resources_Provided}

\section{Experimental Protocol}
\label{sec:experimental_protocol}
\input{3_Experimental_Protocol}

\section{Metrics Adopted}
\label{sec:metrics}

\input{4_Metrics}

\section{Biometric Verification Systems}
\label{sec:biometric_verification_systems}
\input{5_Biometric_Verification_Systems}

\section{Experimental Results}
\label{sec:experimental_results}
\input{6_Experimental_Results}

\section{Limitations}
\label{sec:limitations}
\input{6a_Limitations}

\section{Conclusions and Future Work}
\label{sec:conclusions}
\input{7_Conclusions}


\section*{Acknowledgments}
This project has received funding from the Project INTER-ACTION (PID2021-126521OB-I00 MICINN/FEDER), the European Union’s Horizon 2020 research and innovation programme under the Marie Skłodowska-Curie grant agreement No. 860315 (PriMa project), HumanCAIC project (TED2021-131787BI00 MICINN), and from the Autonomous Community of Madrid (ELLIS Unit Madrid).

\bibliographystyle{unsrt}
\bibliography{0_Main}

\end{document}

%% file: 1_Introduction.tex
Keystroke Dynamics (KD) refers to the typing behavior of human subjects. It is commonly regarded as a \textit{behavioral biometric} trait, similarly to gait \cite{delgado2023exploring}, touch gestures \cite{stragapede2022ijcb, stragapede2023behavepassdb}, signature \cite{TOLOSANA2022108609}, etc. 
In its simplest form, keystroke dynamics are captured as discrete time instants: the time instants a key is pressed and released (for instance in Unix time format), accompanied by the code (ASCII) of the key pressed. Other information such as the amount of pressure on the key or the size of the fingertip might be available depending on the specific hardware capabilities. Consequently, the deployment of systems based on keystroke dynamics is generally economic, as there is no need for specific hardware other than keyboards, which currently represent the primary means for entering textual data into digital systems, with billions of users typing every day. 

Current applications and research directions for KD include: 
\begin{itemize}
    \item[\textit{(i)}] security (biometric recognition), i.e., subject identities can be verified while they write an email or they take a test in educational platforms, or as an additional biometric security layer on top of a traditional knowledge-based password 
    , etc. In comparison with its physiological counterparts such as face or fingerprint, behavioral biometrics represent a more challenging technical problem in terms of recognition performance as they are in general characterized by a higher \textit{intra-user} variability, and lower \textit{inter-user} variability. Nevertheless, behavioral biometrics can offer the advantage of operating \textit{transparently}, without requiring users to complete any additional authentication procedure.
    \item[\textit{(ii)}] forensics: multiple unrelated accounts used by the same subject can be linked by matching their typing behavior, allowing the identification or shortlisting of malicious users. Such measures could help in contrasting toxicity, hate, and harassment on social networks \cite{mandryk2023combating}, protecting minors from online grooming \cite{BORJ2023110039}, preventing the spread of fake news \cite{morales2020keystroke} as well as ``Wikipedia wars'' \cite{haaretz}.    
    \item[\textit{(iii)}] privacy protection: in biometrics, specific patterns in the input data are often associated with demographic groups (such as gender, age, ethnicity). Such patterns can create biases in the biometric systems \cite{2021_TTS_Biases_Terhorst}, leading to worse decisions that affect specific demographic groups, impacting the demographic fairness of the systems. Furthermore, such categories correspond to sensitive attributes, and their disclosure represents a privacy risk. Consequently, the presence of demographic bias often implies that sensitive attributes are embedded in the learned representations, creating the risk of soft-biometric information leakage. 
    The existence of biological differences between different genders, ages, or ethnic groups can be a trivial hypothesis for some biometric characteristics like face, whereas it is not straightforward to make similar assumptions for others. Concerning KD, among others, several studies have evaluated the predictability of gender \cite{8966639}, age \cite{10.1007/978-3-319-91189-2_33}, and even emotions \cite{doi:10.1080/0144929X.2014.907343} and mother tongue \cite{telecom4030021}, laying the foundations for further investigation of KD-based biometric systems from the perspectives of fairness and privacy.  
\end{itemize}


In this scenario, we present the Keystroke Verification Challenge - onGoing (KVC-onGoing), a novel experimental framework to benchmark KD for biometric verification. The framework is hosted on CodaLab\footnote{\href{https://codalab.lisn.upsaclay.fr/competitions/14063/}{https://codalab.lisn.upsaclay.fr/competitions/14063/}}.

The novel contributions introduced by the KVC-onGoing can be listed as follows: 
\begin{itemize}
\item The proposal of a standard experimental protocol and benchmark for biometric verification based on KD, designed to be easily used by the research community. In fact, so far different systems proposed in the literature are generally evaluated in different conditions (datasets, amount of subjects, number of enrolment sessions, and metrics), hindering direct comparisons. The data we provide is structured into a development set (with labels) and evaluation set with a comparison file (Sec. \ref{sec:experimental_protocol}). 
\item The analysis of two different acquisition device scenarios, i.e., desktop and mobile keyboards, each corresponding to a task of the challenge. 
\item The unprecedented size of the datasets, which count on over 185,000 subjects overall (Sec. \ref{sec:resources}). We consider the two largest public databases of keystroke dynamics up to date, the Aalto Desktop \cite{Dhakal2018} and Mobile \cite{palin2019people} Keystroke Databases. Such databases were collected for Human-Computer Interaction (HCI) research, and, although suitable for biometric research, they were not accompanied by any experimental protocol for biometrics purposes (i.e., a list of comparisons). We extracted specific datasets that guarantee a minimum amount of data per subject, age and gender annotations, absence of corrupted data, and that avoid too unbalanced subject distributions with respect to the considered demographic attributes. Previous KD verification systems have mostly been only evaluated with up to several hundred subjects, not representing well the recent challenges that massive usage applications can face, and not considering demographic attributes.
\item The \textit{transcript} text format, in which subjects are asked to read, memorize, and type a text that is presented to them and that is not the same across different samples: consequently, data are much sparser, more unstructured, and they present a higher rate of typing errors, compared to the traditional fixed-text scenario benchmarks \cite{kboc}, which aims to represent for instance the case of an intruder typing the password of the victim. 
\item The completeness of the metrics adopted. The CodaLab platform returns results computed according to several metrics (Sec. \ref{sec:metrics}) that quantify the verification performance as well as the fairness of biometric systems.      
\item A new, clear state-of-the-art performance level. The results on the evaluation set of KVC-onGoing have proved the high discriminative power of keystroke dynamics, reaching values as low as 3.33\% of Equal Error Rate (EER) and 11.96\% of False Non-Match Rate (FNMR) @1\% False Match Rate (FMR) in the desktop scenario, and 3.61\% of EER and 17.44\% of FNMR @1\% at FMR in the mobile scenario, significantly improving previous state-of-the-art results \cite{stragapede2023kvc}. Moreover, several neural architectures are compared, many of them made available by their authors (Sec. \ref{sec:biometric_verification_systems}).
\item The ongoing aspect of the challenge, which allows research teams around the world to benchmark their systems anytime and consequently advance the state of the art of keystroke verification. 
\end{itemize}

A preliminary version of this article was published in \cite{bigdata}, describing the limited-time challenge hosted at the 2023 IEEE International Conference on Big Data (IEEE BigData 2023), in which state-of-the-art results were achieved, setting a high performance level for future work to build on. 
This article significantly improves \cite{bigdata} in the following aspects: \textit{(i)} a major description of the motivation and the experimental protocol proposed in the challenge; \textit{(ii)} a more extensive description of the KD-based verification systems is provided, including key figures illustrating the system architectures and features extracted; \textit{(iii)} we adapt and benchmark our recent systems TypeNet \cite{typenet} and TypeFormer \cite{typeformer} to serve as state-of-the-art baseline methods for our comparative experiments; \textit{(iv)} incorporating additional metrics in the evaluation of the proposed systems in order to analyze different operational scenarios; \textit{(v)} an in-depth analysis is carried out based on the results according to different biometric operating points; \textit{(vi)} all systems are analyzed from the point of view of fairness.

The remainder of the article is organized as follows: first, related work is briefly introduced (\ref{sec:related_work}), followed by a presentation of the resources provided within the proposed experimental framework(Sec. \ref{sec:resources}). Then, Sec. \ref{sec:experimental_protocol} includes a detailed presentation of the evaluation protocol of the experimental framework and challenge. Sec. \ref{sec:metrics} presents the metrics adopted. Sec. \ref{sec:biometric_verification_systems} provides a thorough descriptions of the biometric systems benchmarked so far. Finally, Sec. \ref{sec:experimental_results}, Sec, \ref{sec:limitations}, and Sec. \ref{sec:conclusions} respectively contain the analysis of the results obtained, of their limitations, and the article conclusive remarks.

%% file: 2a_Related_Work.tex
Classifying subjects by their typing behavior began with the rise of personal computers. Keystroke biometrics started with mechanical keyboards, being the only available hardware, but later touchscreen devices have also been incorporated in the scope of KD research. The advancements in terms of processing power and the machine learning algorithms have not spared the field of KD, leading to significant improvements in the biometric recognition performance over the years. To this end, it is well known that the scale of training datasets plays a pivotal role in boosting the performance of such models. Table \ref{tab:old_dbs} features some of the most important databases for KD, showing that the size of the Aalto Keystroke Databases is significantly greater than other public databases, apart from adopting different text formats. 

\begin{table}[t!]

\centering
\footnotesize
\caption{\small Some of the most important public keystroke dynamics databases in chronological order. D stands for desktop, M stands for mobile.}
\begin{tabular}{c|c|c|c|c} 
\textbf{Database} & \textbf{Scenario} & \makecell{\textbf{No. of}\\ \textbf{Subjects}} & \textbf{Text Format} & \makecell{\textbf{Strokes}\\ \textbf{per Subject}} \\ 
\hline
\makecell{GREYC\\ (2009) \cite{greyckeytroke}} & D & 133 & Fixed & $\sim$800 \\
\makecell{CMU\\ (2009) \cite{cmu}} & D & 51 & Fixed & $\sim$400\\ 
\makecell{BiosecurID\\ (2010) \cite{biosecurid}} & D & 400 & Free & $\sim$200 \\
\makecell{RHU\\ (2014) \cite{rhu}} & D & 53 & Fixed & $\sim$600 \\ 
\makecell{Clarkson I\\ (2014) \cite{clarksonI}} & D & 39 & Fixed, free & $\sim$20k\\
\makecell{SUNY\\ (2016) \cite{sun2016shared}} & D & 157 & Transcript, free & \makecell{$\sim$17k} \\ 
\makecell{Clarkson II\\ (2017) \cite{clarksonII}} & D & 103 & Free & $\sim$125k \\
\makecell{\textbf{Aalto Desktop}\\ \textbf{(2018) \cite{Dhakal2018}}} & \textbf{D} & \textbf{168k} & \textbf{Transcript} & \makecell{$\sim$\textbf{750}}\\
\makecell{\textbf{Aalto Mobile}\\ \textbf{(2019) \cite{palin2019people}}} & \textbf{M} & \textbf{37k} & \textbf{Transcript} & $\sim$\textbf{750}\\
\makecell{HuMIdb\\ (2020) \cite{acien2021becaptcha}} & M & 600 & Fixed & \makecell{$\sim$20} \\
\makecell{BehavePassDB\\ (2022) \cite{stragapede2023behavepassdb}} & M & 81 & Free & \makecell{$\sim$100} \\
 \end{tabular}
 \label{tab:old_dbs}
\end{table}

Concerning model architectures, among the first studies which benchmarked neural networks against traditional machine learning and hand-crafted algorithms is \cite{cmu}. Then, convolutional \cite{ceker2017} and recurrent neural networks \cite{sun2016shared} have been employed to extract higher-level keystroke features. In \cite{Li2022}, keystroke sequences are structured as image-like matrices and processed by a CNN combined with a Gated Recurrent Unit (GRU) network.

The Aalto Keystroke Databases were adopted to train and evaluate TypeNet \cite{typenet}, a Long Short-Term Memory (LSTM) Recurrent Neural Networks (RNN) trained with triplet loss, proposed by Acien \textit{et al.} This was an important milestone in KD research at large scale, compared to previous approaches using small and limited databases. In \cite{morales2022setmargin}, a novel loss function, called SetMargin loss (SM-L), was proposed to address the challenges associated with keystroke biometrics in which the classes used in learning and inference are disjoint. This led to improved recognition perfomance with TypeNet.

Later on, replicating the same experimental protocol used in \cite{typenet}, Stragapede \textit{et al.} proposed TypeFormer \cite{typeformer}. TypeFormer was developed starting from the Transformer model \cite{vaswani2017attention}, with several adaptations to optimize its recognition performance for KD. In several experiments, TypeFormer outperformed TypeNet in the mobile environment, but not in the desktop case \cite{stragapede2023kvc}.

Recently, in \cite{neacsu2023doublestrokenet}, a novel approach called DoubleStrokeNet for recognizing subjects using bigram embeddings was proposed. This is achieved using a Transformer-based neural network that distinguishes between different bigrams. The authors experimented with the Aalto Keystroke Databases, reaching interesting results in terms of recognition performance, although using a different experimental protocol as in \cite{typenet, typeformer, stragapede2023kvc}.

\begin{table*}[!t]
\centering
\caption{\small Demographic distributions of the provided datasets. The rows represent different age groups, while the columns represent genders. The evaluation sets are balanced with respect to gender. The notations ``Labelled'' of ``Unlabelled'' refer to the number of subjects for whom the age and gender labels are or are not available.}
\begin{tabular}{c|cc|cc|cc|cc} 
\toprule
& \multicolumn{4}{c|}{\textbf{Desktop Task}} & \multicolumn{4}{c}{\textbf{Mobile Task}} \\
\bottomrule
& \multicolumn{2}{c|}{\textbf{Development Set}} & \multicolumn{2}{c|}{\textbf{Evaluation Set}} & \multicolumn{2}{c|}{\textbf{Development Set}} & \multicolumn{2}{c}{\textbf{Evaluation Set}} \\ 
\bottomrule
\textbf{Age Group} & \textbf{Male} & \textbf{Female} & \textbf{Male} & \textbf{Female} & \textbf{Male} & \textbf{Female} & \textbf{Male} & \textbf{Female}\\   
\hline
 \textbf{10 - 13} & 4,336 & 5,420 & 1,085 & 1,085 & 622 & 800 & 254 & 254 \\ 
 \textbf{14 - 17} & 10,993 & 8,336 & 1,861 & 1,861 & 1,537 & 1,516 & 618 & 618 \\ 
 \textbf{18 - 26} & 25,752 & 24,315 & 1,861 & 1,861 & 4,359 & 8,999 & 843 & 843 \\ 
 \textbf{27 - 35} & 9,607 & 12,281  & 1,861 & 1,861 & 1,343 & 4,002 & 547 & 547 \\  
 \textbf{36 - 44} & 2,143 & 5,331 & 536 & 536 & 382 & 1,333 & 156 & 156 \\ 
 \textbf{45 - 79} & 1,182 & 5,424 & 296 & 296 & 200 & 739 & 82 & 82 \\
\bottomrule
\end{tabular}\\
Desktop task: Labeled: 115,120 (Development), 15,000 (Evaluation), Unlabeled: 0 (Development, Evaluation)\\
Mobile task: Labeled: 25,832 (Development), 5,000 (Evaluation), Unlabeled: 14,807 (Development), 0 (Evaluation)\\
Although unlabeled, we opted to include some subjects in the mobile task development set to maximize its size.\\

\label{table:demo}
\end{table*}

%% file: 2_Resources_Provided.tex
The proposed experimental framework is based on the two most complete and large-scale public databases of KD up to date, collected by the User Interfaces\footnote{\href{https://userinterfaces.aalto.fi/}{https://userinterfaces.aalto.fi/}} group of the Aalto University (Finland). The two databases are collected respectively in a desktop\footnote{\href{https://userinterfaces.aalto.fi/136Mkeystrokes/}{https://userinterfaces.aalto.fi/136Mkeystrokes/}} \cite{Dhakal2018} and mobile\footnote{\href{https://userinterfaces.aalto.fi/typing37k/}{https://userinterfaces.aalto.fi/typing37k/}} \cite{palin2019people} acquisition environment, including respectively around 168,000 and 60,000 subjects.

Such large-scale data collections were promoted by their authors with the goal of analyzing HCI performance, through statistical analyses on the correlation between typing performance (words per minute, error rates, difference between hands, etc.) and various factors, such as demographics, finger usage, and use of intelligent text entry techniques, etc. Nonetheless, the suitability of the collected data for research on biometric recognition was also highlighted by the authors.

The data collection was launched globally in collaboration with a commercial organization offering online typing courses and tests\footnote{\url{https://www.typingmaster.com/}}. Data were collected in a web-based transcription task hosted on a university server, following the guidelines of online study platforms \cite{reinecke2015labinthewild}. 

The transcription task was designed according to a common practice in text entry research, known as the \textit{unconstrained text entry evaluation paradigm} \cite{wobbrock2007measures}.
Each of the acquisition sessions contains a sentence of transcript text (variable content, but not fully free-text) captured in an unsupervised way. Subjects were asked to read, memorize, and type in their device English sentences that were randomly selected from a set of 1,525 sentences. To distinguish the device type, HTML requests to the site were detected as either \textit{desktop} or \textit{mobile}. Data coming from a single subject were recorded from at once, therefore with a single acquisition device per subject. No requirements on a minimum time gap between consecutive tasks were posed. The subjects were asked to write 15 sentences consecutively, however, not all subjects completed the task. For the KVC challenge we only selected subjects who had completed all the 15 sentences.  

When all sentences had been transcribed, participants had to fill in a demographic questionnaire. It asked their age and gender, country and native language, keyboard type and layout, typing experience, and number of fingers used for typing. In addition, system information such as the device model, the browser used, and the screen resolution was saved for each session and it is available alongside the raw data.

The raw data acquired consist of the timestamp of the instant a key is pressed, the timestamp of the instant the key is released, and the key ASCII code. After discarding some of the subject data due to insufficient acquisition sessions per subject (less than 15 per subject), within the scope of this work, the two databases as downloaded have been rearranged to form four datasets,  for both development and evaluation of keystroke-based biometric verification system. 

Table \ref{table:demo} shows the division into development and evaluation datasets for both tasks, as well as demographic distribution of the datasets provided in the competition. The subjects have been divided into six age groups (10 - 13, 14 - 17, 18 - 26, 27 - 35, 36 - 44, 45 - 79). A validation set is not explicitly provided, but it can be obtained from the development set according to different training approaches. The data are accompanied by the respective lists of pairwise comparisons to be carried out with two Python script files to load the data, and run the comparisons, generating a text file ready to be submitted to CodaLab for scoring.

%% file: 3_Experimental_Protocol.tex

The two tasks (desktop and mobile) are structured similarly, and they are designed for a biometric verification protocol. A biometric sample corresponds to an acquisition session. For the desktop task, 2,250,000 1vs1 session-level comparisons are carried out, involving 15,000 subjects not included in the development set. For the mobile task, 750,000 comparisons, involving 5,000 subjects not included in the development set. The proposed experimental framework follows an open-set learning protocol, in other words, the subjects in the development and evaluation sets are different.


For each subject, there are 5 enrolment sessions and 10 verification sessions, leading to 50 1vs1 comparisons, which are averaged over the 5 enrolment sessions generating 10 genuine scores per subject. In a similar manner, 20 impostor scores per subject are generated. The impostor sessions are divided into two groups: 10 \textit{similar} impostor scores, for which the verification sessions are randomly selected from subjects belonging to the same demographic group (same gender and age); 10 \textit{dissimilar} impostor scores, in which the verification sessions are all randomly selected from subjects of different gender and age intervals. 

Based on the described evaluation design, following \cite{typenet, typeformer}, we consider two cases for evaluating the system: \textit{(i)} global distributions: this case corresponds to dividing all scores into two groups, genuine and impostor scores, regardless of which subject they belong too. This case corresponds to having a fixed, pre-determined threshold, implying a simpler deployment of the biometric system. In order to assess the performance of the biometric system, this choice means setting one single threshold for all comparisons to obtain a decision. \textit{(ii)} mean per-subject distributions: the optimal threshold is computed at subject-level, considering the 30 verification scores as described above. This choice corresponds to providing the system with more flexibility, so that it can adapt to user-specific distributions \cite{typenet}. 

%% file: 4_Metrics.tex

To provide a complete assessment, the CodaLab platform returns results computed according to the several metrics reported below.

For biometric performance, the metrics considered are: Equal Error Rate (\%) (global and mean per-subject) based on which the competition ranking is created, False Non-Match Rate at X\% False Match Rate (FNMR @ X\% FMR) considering X = 0.1\%, 1\%, 10\% (global), Area Under the ROC Curve (AUC) (global and mean per-subject), accuracy (global and mean per-subject), rank-n, n=1 (mean per-subject). Some of the metrics are not computed considering the mean per-subject distributions as there are not enough scores per subject for a sufficiently fine threshold resolution. In the case of rank-n, it is only possible to compute it at subject level. This metric concerns the identification of subjects (i.e., 1 to many comparisons), therefore assessing a different scenario from the previous metrics, which refer to the case of verification (i.e., binary classification). Additionally, the Detection Error Tradeoff (DET) curve, the Receiving Operator Characteristic (ROC) curve, and the global distribution histograms are plotted. 

For quantifying fairness, the metrics considered are: Standard Deviation of global verification accuracy by demographic group (STD), Skewed Error Rate of global verification accuracy by demographic group (SER), Fairness Discrepancy Rate (FDR) \cite{9507539}, Inequity Rate \cite{grother2019face}, Gini Aggregation Rate for Biometric Equitability (GARBE) \cite{howard2022evaluating}, and Skewed Impostor Rate (SIR) \cite{stragapede2023kvc}. A complete explanation of all the mentioned metrics is provided in \cite{stragapede2023kvc}. All functions used to compute the mentioned metrics are made available\footnote{\href{https://github.com/BiDAlab/Keystroke_Verification_Challenge}{https://github.com/BiDAlab/Keystroke\_Verification\_Challenge}}.

%% file: 5_Biometric_Verification_Systems.tex
\begin{table*}[t!]
\caption{\small High-level comparison of the proposed keystroke biometric verification systems. \textit{D} stands for desktop, \textit{M} stands for mobile.}
\centering
\begin{tabular}{c|c|c|c|c|c|c}
 \cmidrule[.75pt]{4-7}
\multicolumn{3}{c}{} & \multicolumn{2}{c|}{\textbf{Desktop}} & \multicolumn{2}{c}{\textbf{Mobile}}  \\ 
\bottomrule
\makecell[c]{\textbf{System}} & \makecell[c]{\textbf{Architecture}} & \makecell[c]{\textbf{Loss Function}} & \textbf{Position} & \makecell[c]{\textbf{Global}\\ \textbf{EER (\%)}} & \makecell[c]{\textbf{Position}} & \makecell[c]{\textbf{Global}\\ \textbf{EER (\%)}} \\
\bottomrule
LSIA & CNN+RNN with attention mechanism & Custom & 1 & 3.33  & 1 & 3.61 \\
VeriKVC & CNN & ArcFace & 2 & 4.03 & 2 & 3.78 \\
Keystroke Wizards & GRU (\textit{D}), Transformer (\textit{M}) & Triplet & 3 & 5.22 & 5 & 5.83 \\
U-CRISPER & GRU-Based Siamese Network & Triplet & 4 & 6.19 & 6 & 8.76 \\
YYama & Transformer+CNN & Contrastive + Cross-entropy & 5 & 6.41 & 3 & 4.16 \\
BiDA Lab & RNN (\textit{D}), Transformer (\textit{M}) & Triplet & 6 & 6.76 & 7 & 9.45 \\
Challenger$\ast$ & Transformer & Triplet & 7 & 6.79 & 4 & 5.19 \\
BioSense & CNN with attention mechanism & Cross-entropy & 8 & 10.85 & 8 & 11.83 \\
\bottomrule
\multicolumn{7}{l}{Please see footnote 14.}
\end{tabular}
\label{tab:archi_comp}
\vspace{0.2cm}
\end{table*}

A summary of the biometric systems is proposed in Table \ref{tab:archi_comp}. The challenge ranking is based on the global EER (\%).

\subsection{Team: LSIA}

The LSIA team is composed of one member from Laboratorio de Sistemas de Informaci\'on Avanzados (LSIA), Universidad de Buenos Aires (Argentina).

\paragraph{Model Architecture}

\begin{figure*}[t!]
\includegraphics[width=\linewidth]{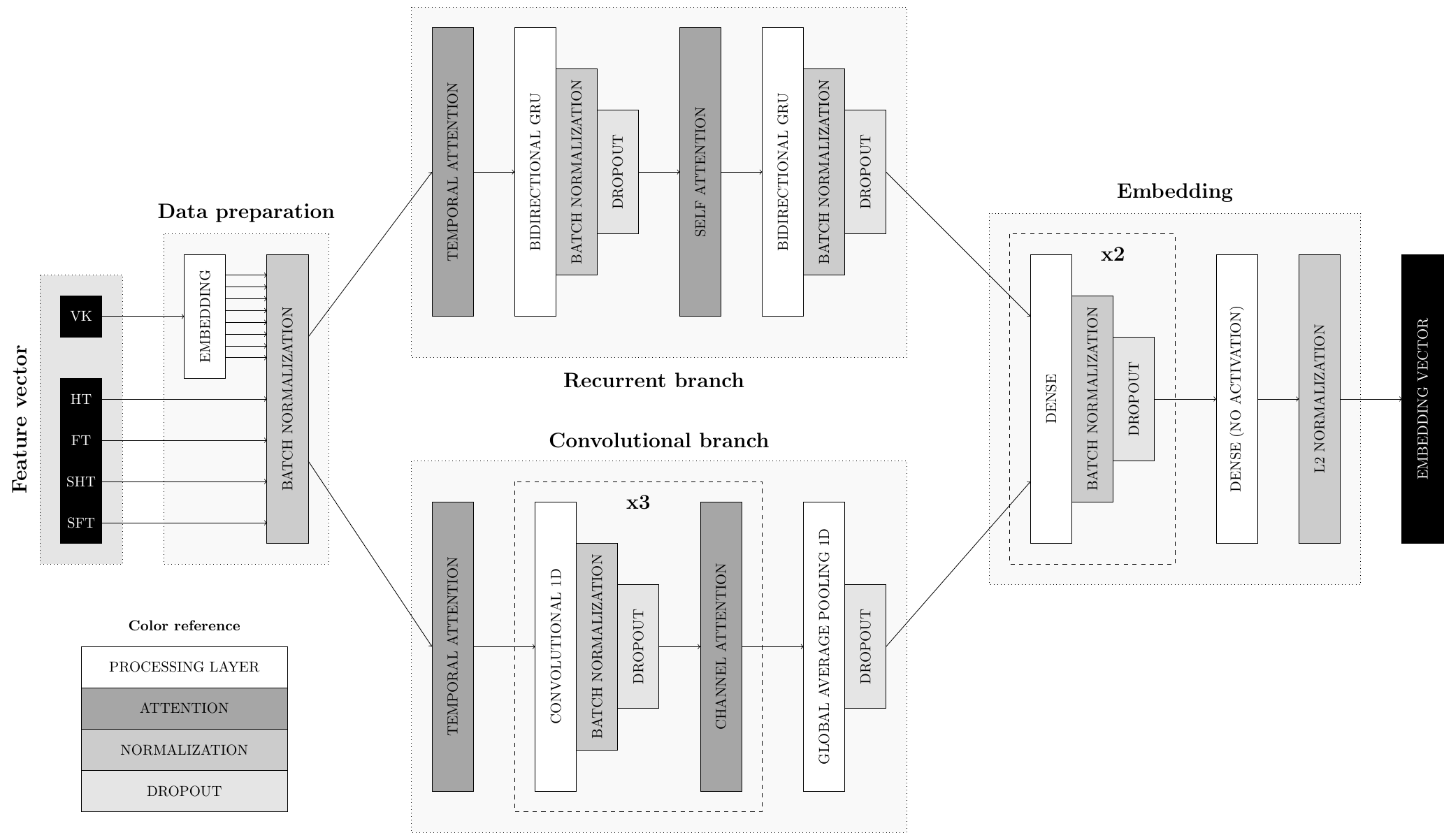}
\caption{\small The dual-branch (recurrent and convolutional) embedding model for distance metric learning proposed by the LSIA team, ranked first in both tasks.}
\label{fig:LSIA}
\end{figure*}

Among the architectures proposed by the LSIA team, the best performance in terms of EER is achieved by a dual-branch (recurrent and convolutional) embedding model for distance metric learning, depicted in Fig. \ref{fig:LSIA}. The same architecture is adopted for both tasks. 

The recurrent branch comprises two bidirectional GRU layers (512 units), while the convolutional branch features three blocks of 1D convolution, each with an increasing number of filters (256, 512, and 1024, with kernel size equal to 6) and utilizes global average pooling. Temporal attention serves as the first layer of both branches. Scaled dot-product attention is applied between the recurrent layers, and channel attention follows each convolutional layer. The outputs of both branches are concatenated, and the final embedding vector is produced by three dense layers. Batch normalization and dropout are used after each processing layer, in both branches and the embedding module, except for the last dense layer, which does not use activation and is followed by L2 normalization as is common for embedding models. 
Sample similarity is measured, as is customary in distance metric learning, using the Euclidean distance between their respective embeddings. 

The choice of architecture is motivated by the observation that keystroke timings result from a combination of two factors: a partially conscious decision process involving what to type and an entirely unconscious motor process pertaining to how to type \cite{gonzalez2021shape}. The convolutional branch is expected to excel at identifying common, short sequences, while the recurrent branch is expected to capture the user’s time-dependent decision process. Empirical testing confirms that a dual-branch model outperforms a purely recurrent or purely convolutional model with an equivalent number of trainable parameters.

\paragraph{Training}

For training the model, only the development sets provided by the organizers are used. These sets are preprocessed to generate five attributes: the integer ASCII key code (VK), for which the network learns a small-dimensional embedding, and four normalized timing features, using a scale of seconds. These are the interval between key press and release events, called the Hold Time (HT), the latency between successive key press events, called Inter-Press Time (IPT), also known as Flight Time (FT), and two synthetic features (SHT and SFT) meant to capture variations in the user's typing style compared to the general population \cite{gonzalez2022towards}. 

A novel loss function is used to minimize the EER of a keystroke dynamics verification system using one-shot evaluation with a uniform global detection threshold across all users. This function extends the SetMargin loss of Morales \textit{et al.} \cite{2022_PR_SetMargin_Morales} to sets of sets (rather than pairs of sets) and includes an additive penalty term to encourage the model to embed all classes within hyperspheres with similar average radii. 

A learning curriculum of increasing difficulty is adopted to train the model. Each batch consisted of 15 samples from 40 different users, totaling 600 samples per batch. In each batch of epoch \textit{i}, the first user is paired with its \textit{i} nearest neighbors based on the proximity of its embedding centers.


\subsection{Team: VeriKVC}

The VeriKVC team is composed of two members from two different companies, respectively Verigram LLC (Singapore), and Citix (Kazakhstan). The code of the described solution is available online\footnote{\href{https://gitlab.com/vuvko/kvc}{https://gitlab.com/vuvko/kvc}}.

\paragraph{Model Architecture}
The system architecture proposed by the VeriKVC team relies on a Convolutional Neural Network (CNN) design, due to its capability to effectively capture spatial features in web behavior data. The CNN model is tailored to process session data and extract significant patterns and features. The same architecture is adopted for both tasks. The CNN has 128 starting features, and an embedding size of 512. The sequence length considered is 128.

In addition to the chosen architecture, feature engineering techniques are employed to preprocess the behavior data. One such technique involves encoding the timing differences of events using both sine and cosine functions \cite{vaswani2017attention}. This normalization helps to account for the varying time scales of events and maintain the relevance of timing information. Furthermore, the difference between the end and start times of events is processed using cosine functions to scale it within the range of -1 to 1. These features extracted by the attention mechanism are essential in providing the model with a better understanding of the temporal aspects of web behavior data.

In response to the potential risk of overfitting, various randomization techniques are utilized for a data augmentation process during training. These techniques introduce variability into the training data, preventing the model from memorizing the training set and promoting its generalization to unseen data. By enhancing the model robustness through data augmentation, its ability to handle diverse real-world scenarios effectively is ensured.

\paragraph{Training}

To train the model, the number of epochs was set to 5000 with a batch size of 512. This extended training duration allows the model to learn complex patterns and improve its overall performance. The AdamW optimizer 
with learning rate of 0.01 and gradient clipping and cosine annealing scheduler 
is chosen. This allows the network to learn basic features quickly and spend more time on tuning more subtle features. In terms of loss function, ArcFace \cite{deng2019arcface} is selected as the most effective choice for the model. This loss encourages the model to learn distinct feature representations for each individual, which is crucial for subject verification tasks. 

\subsection{Team: Keystroke Wizards}

The Keystroke Wizards team is composed of members of the Regensburg University of Applied Sciences (Germany). This is one of the two teams from the same institution\footnote{Team members: Leon Schmitt, Daniel Wimmer, Christoph Großmann, Joerdis Krieger, and Florian Heinz.}. 

\paragraph{Model Architecture}

In the desktop task, the proposed model is based on a RNN combined with a triplet loss function, following the approach in \cite{typenet}. In contrast, in the mobile task, the proposed model is based upon the models proposed in \cite{typeformer}, and \cite{senerath2023behaveformer}, which all use Gaussian Range Encoding (GRE) as well as triplet loss.

The desktop model employs a Multi-Layer GRU RNN with 11 normalized input features: HT, HT of the following key, IPT, time from first key press to following key release, Inter-Key Time (IKT), Inter-Release Time (IRT), rollover duration, rollover count, hold-to-rollover ratio, ASCII code, and ASCII code of following key. The rollover-duration feature describes for how long both keys of the key pair are pressed concurrently. The rollover-count describes how often both keys are pressed at the same time. The hold-to-rollover-ratio describes the ratio of the total hold times to the total rollover times of the session. These features are chosen based on data analysis, correlation analysis, and scatter plots. 

The mobile model is a hybrid Transformer with a Channel Module and a Temporal Module, incorporating Multi-Scale CNN and GRU layers. In addition to desktop features, they use trigram features, capturing the timing of the first and third key presses. The mobile model integrates GRUs for temporal data and CNNs for spatial data. It features multi-head attention and Gaussian range positional encoding. 

\paragraph{Training}

Both models are trained with the Triplet Loss function. The desktop model utilizes a PyTorch-based architecture featuring batch normalization, GRU layers, dropout, and linear layers. Training parameters include 160 epochs, 128 batches per epoch, 1024 sequences per batch, a sequence length of 70, and 80\% data for training. The mobile model is trained with 160 epochs, a sequence length of 50, and 80\% training data.


\subsection{Team: U-CRISPER}
The U-CRISPER team is composed of members from the Regensburg University of Applied Sciences (Germany). This is one of the two teams from the same institutions\footnote{Team members: Ron Krestel, Christoff Mayer, Simon Haberl, Helena Gschrey}. 

\paragraph{Model Architecture}

Four essential time-based features are derived: HT, IKT, IPT, and IRT. These features are standardized by removing the mean and scaling to unit variance. Any missing value is substituted by the mean. The ASCII codes, normalized by dividing them by 255, are also used. The raw data also contain upward and downward outliers, which could potentially introduce confusion into the model. To address this issue, an in-depth analysis of the data point distribution is conducted. As a result, extreme outliers are clipped to a predefined boundary, enhancing the model's generalization.
The adopted model architecture is a GRU-Based Siamese Network, utilizing GRUs to adaptively capture temporal dependencies. Therefore, the design fits well for handling keystroke sequences. Two GRU layers, each with a hidden size of 64 and a dropout rate of 0.2 in between, process an input sequence (70 by 5) and a Linear Layer with a prior Rectified Linear Unit (ReLU) activation generates an embedding vector of dimension 64.

\paragraph{Training}

For effective training, the Triplet Loss function \cite{typenet} is used. Each comparison is carried out at session level: one session is used as an anchor, one as a positive from the same user, and one as a negative from a different user. The model processes each session, generating embeddings compared by the Triplet Loss function, which minimizes the anchor-positive distance and maximizes the anchor-negative distance.

An hyperparameter optimization is conducted by using Optuna's Tree-structured Parzen Estimator (TPE) \cite{watanabe2023tree}, resulting in the following model configuration: 200 epochs, 70-length input sequences,  batch size of 512 (150 for training, 40 for validation), Adam optimizer (LR: 0.001, betas: 0.9, 0.999, epsilon: 1e-08), and Triplet Loss (margin: 1.5, p-value: 2.0).


\subsection{Team: YYama}
The YYama team is composed of one member from the Division of Radiology and Biomedical Engineering, The University of Tokyo (Japan). The code of the solution described is available online\footnote{\href{https://github.com/yamagishi0824/kvc-dualnet}{https://github.com/yamagishi0824/kvc-dualnet}}.

\paragraph{Model Architecture}

The proposed ``dual-network'' approach combines an embedding model for feature extraction and a classifier model for verification, aiming to improve the biometric verification performance. 


The development dataset is randomly split into 80\% for training and 20\% for validation. The feature extraction stage included five conventional variables (HT, IKT, IPT, IRT, and ASCII code) \cite{typenet}, \cite{typeformer}, plus two new variables (differential in consecutive HTs and ASCII codes). These features are standardized and normalized to maintain consistency across the dataset.

The same architecture is used for both tasks. The embedding model architecture is a simplified transformer-based design featuring 1D CNN layers followed by a transformer encoder and a max pooling operation, concluding with a fully connected layer. In contrast, the classifier model is a straightforward 1D CNN-based architecture that took paired outputs from the embedding model to determine if they originate from the same user.

\paragraph{Training}
The training settings are consistent across both tasks, with the embedding model using the Adam optimizer and a fixed learning rate of 0.001, trained for 2,000 epochs for mobile and 1,000 epochs for desktop. The batch size during training is 2,048, with a total of 16 batches supplied to the models per epoch. Therefore, the total number of data pairs per epoch was 32,768. The loss function is contrastive loss with a margin of 1 for the embedding model. Simultaneously, the classifier model's embeddings are trained using the binary cross-entropy loss function with the same learning rate as the embedding model.

Model performance is evaluated using the global EER, and the best models for mobile and desktop are selected based on the lowest EER. The dual-network approach yields lower EERs than conventional approaches based on Euclidean distance.



\subsection{Team: BiDA-Lab}
The BiDA-Lab team is composed of members from the Universidad Autonoma de Madrid (Spain). The results reported correspond to the initial benchmark carried out by the organizers of the KVC \cite{stragapede2023kvc}.

\paragraph{Model Architecture}

Throughout the proposed framework, we evaluate two recent state-of-the-art deep-learning models:
\begin{itemize}
    \item TypeNet \cite{typenet}: a LSTM RNN, trained with triplet loss. In this case, we consider input sequences of 150 characters typed. TypeNet is implemented in Tensorflow.
    \item TypeFormer \cite{typeformer}: a novel transformer architecture consisting in a temporal and a channel module enclosing two LSTM RNN layers, a Gaussian Range Encoding (GRE), a multi-head self-attention mechanism, and a block-recurrent transformer structure. TypeFormer is also trained with triplet loss. In this case, we consider input sequences of 50 characters typed. TypeFormer is implemented in PyTorch and publicly available\footnote{\url{https://github.com/gstrag/TypeFormer}}.
\end{itemize}
Both approaches utilize distance metric learning \cite{2022_PR_SetMargin_Morales}: the models are trained to transform input data into a new feature space, enabling straightforward distances to be used for analyzing and leveraging the ``semantic'' arrangement of the input space. The Euclidean distance is chosen and the Triplet loss is used to train the model. 
For evaluation, the distances obtained from the comparisons of feature embeddings corresponding to each of the test sessions are normalized, and then they are subtracted from 1 in order to transform them into similarity scores.

\paragraph{Training}
The training of both models takes place on the challenge development set considering identical settings to those described in their respective papers. The only differences are related to the division into training and validation sets. For TypeNet, we consider a subset of 400 subjects to validate the model at the end of each training epoch in terms of average EER per subject. This choice is justified by the experimental protocol followed in \cite{typenet}. For TypeFormer, we consider an 80\%-20\% train-validation division of the KVC-onGoing development set, and we adopt the global EER as validation metric. According to these validation metrics, the best-performing epoch model is saved in each case.

\subsection{Team: Challenger}
The Challenger team is composed of three members from the National University of Singapore, and one member from the University of Moratuwa (Sri Lanka).

\paragraph{Model Architecture}
Starting from the raw data, some features are extracted considering digram and trigram features. The time intervals between various events to create features are also used. The combination of these time intervals and ASCII codes yields a total of 10 keystroke features. ASCII code is normalized to the range [0,1] and time based features are represented in seconds.

A Spatio-Temporal Dual Attention \cite{senerath2023behaveformer} method for improved attention and feature extraction across time and modalities is employed. The proposed model is based on a variation of the Vanilla Transformer encoder named Spatio-Temporal Dual Attention Transformer, that is based on a framework named BehaveFormer \cite{vaswani2017attention, senerath2023behaveformer}. In this framework, the raw data are preprocessed to extract additional features. Then, the Transformer architecture with Dual Attention modules are used to extract more discriminative features from KD. Then, the Spatio-Temporal Dual Attention Transformer uses a single transformer with two attention mechanisms: one over the temporal axis and one over the channel axis. This allows the model to focus on the relevant keystroke features over time, extracting unique behavioral patterns for individual users. The GRE is used as the positional encoding. Moreover, it uses a 2D convolution network instead of the feed forward network in the transformer blocks, to analyze the input data over time and across different channels.


\paragraph{Training}

To give the model a better starting point than the randomly assigned weights, an initial model is pre-trained which is a good generalized starting point for the keystroke verification task, providing a solid base for the further training on the data provided by the organizers\footnote{Having used the raw data of the Aalto Databases, from which the KVC-onGoing datasets are obtained, it is likely that some of the subjects in the evaluation set were also used for the pre-training, lifting the open set learning protocol restriction, according to which the development and evaluation sets should not have any subject in common. A dedicated scenario, called Unrestricted, is included in the KVC-onGoing accounting for the option of pre-trained models. Being this the only team that participated in this form, it is included in the general rankings, but with a special mark $\ast$.}. For the pre-training, keystroke data from these datasets are used: Aalto Databases \cite{Dhakal2018, palin2019people}, HMOG \cite{yang2014multimodal}, and HuMidb \cite{acien2021becaptcha}. The Triplet Loss function is used for training. Two different models with the same architecture are trained for the two tasks. The models were trained with a batch size of 512 and for 1050 epochs for the desktop model and 2500 epochs for the mobile .

\subsection{Team BioSense}
The BioSense team is composed of members from the NASK – National Research Institute (Poland). The code of the Anabel-KA is available online\footnote{\href{https://github.com/nask-biometrics/anabel-ka}{https://github.com/nask-biometrics/anabel-ka}}.

\paragraph{Model Architecture}

A model of a neural network with Keycode Attention, named Anabel-KA, is developed. Keycode Attention refers to the self-attention \cite{vaswani2017attention} module, in which the attention matrix is computed from the sequence of ASCII codes for the keyboard buttons pressed by the users, and values in the attention function are derived from the time intervals between pressing and releasing keys in the sequence. $Q$ and $K$ (in the equation of scaled dot product attention below) are a result of the keycode extraction function $F_{KB}$ and $V$ is a result of the time-interval extraction function $F_{T}$.

\begin{align*}
\label{eq:attention}
Att(Q,K,V)=&softmax(\frac{QK^T}{\sqrt{d_k}})V\\ 
&\mbox{where }Q=K=F_{KB}(X),V=F_{T}(X) \notag
\end{align*}

Feature extraction $F_{KB}$ from the keycodes is made by a set of 2D convolutions with kernels of size 3, as authors of other methods described in the literature~\cite{gonzalez} show the feasibility of analyzing 3-element groups of keys for the keystroke biometrics. Typing dynamics in the proposed model are represented by the sequence of HTs and IKTs. Such values are processed by the time extractor module $F_{T}$ which is also built using 2D convolutions of size 3 and stride 1. In both cases, the time extractor and the keycode extractor, the number of convolution filters is chosen to be 64. Input sequences are normalized and the length is arbitrarily set to 66, as the median length of the inputs in the KVC-onGoing database is equal to 48 for mobile and desktop entries. Longer sequences are cut and shorter ones are padded with zeros. Extension of the model with the method for handling longer sequences is planned to be added in the future. Embeddings generated by the Anabel-KA have a length of 256. The Keycode Attention is implemented with the multi-head attention module with 8 heads. 

\paragraph{Training}
For training, an additional classification layer is used with a number of outputs equal to the number of classes and softmax function applied. When evaluating, the similarity between keystroke sequences is computed using the Euclidean distance between their embeddings. Training is done with the cross-entropy loss function. For the purpose of the experiments, a cross-validation with 10 folds is employed, with 500 subjects excluded from the dataset for validation in each fold. The maximum number of epochs is set to 60. The final model is selected based on the maximum value of the validation metric, which is the AUC. The batch size is 128 and optimization is done with the Adam algorithm, the training procedure starts from the warm-up learning rate of 0.01 and changes gradually after every 10 epochs. 


\begin{table*}[htbp]
\centering
\caption{\small Comparison of the results achieved in the desktop (top half) and mobile tasks (bottom half) considered in KVC-onGoing. Each half is further split into global distributions (upper part) and mean per-subject distributions (lower part).}
\begin{tabular}{c|c|c|c|c|c|c|c}

\multicolumn{8}{c}{\textbf{Desktop}} \\
\bottomrule
\textbf{Position} & 
\textbf{Team} & 
\makecell{\textbf{EER (\%)}$\downarrow$} &
\makecell{\textbf{FNMR @0.1\%}\\ \textbf{FMR (\%)$\downarrow$}} & 
\makecell{\textbf{FNMR @1\%}\\ \textbf{FMR (\%)$\downarrow$}} & 
\makecell{\textbf{FNMR @10\%}\\ \textbf{FMR (\%)$\downarrow$}} &
\textbf{AUC (\%)$\uparrow$} &
\textbf{Accuracy (\%)$\uparrow$} \\
\bottomrule
1 & LSIA & 3.33 & 44.17 & 11.96 & 0.51 & 99.48 & 96.68 \\
2 & VeriKVC & 4.03 & 59.05 & 18.79 & 1.05 & 99.07 & 95.97 \\
3 & Keystroke Wizards & 5.22 & 67.86 & 27.98 & 1.62 & 98.79 & 94.78 \\
4 & U-CRISPER & 6.19 & 74.77 & 35.24 & 2.68 & 98.37 & 93.81 \\
5 & YYama & 6.41 & 74.16 & 36.96 & 2.88 & 98.28 & 93.59 \\
6 & BiDA Lab & 6.76 & 77.4 & 39.57 & 3.45 & 98.08 & 93.24 \\
7 & Challenger$\ast$ & 6.79 & 76.93 & 39.36 & 3.52 & 98.09 & 93.21 \\
8 & BioSense & 10.85 & 84.94 & 54.59 & 12.0 & 95.86 & 89.15 \\
\bottomrule

\end{tabular}
\vspace{0.2cm}

\begin{tabular}{c|c|c|c|c|c}


\multicolumn{6}{c}{Mean Per-Subject Distributions} \\
\bottomrule
\textbf{Position} & \textbf{Team} & 
\textbf{EER (\%)$\downarrow$} & \makecell{\textbf{AUC}\\  \textbf{(\%)$\uparrow$}} & \makecell{\textbf{Accuracy}\\  \textbf{(\%)$\uparrow$}} & \makecell{\textbf{Rank-1}\\\textbf{(\%)$\uparrow$}} \\
\bottomrule
1 & LSIA & 0.77 & 99.87 & 96.43 & 98.04 \\
2 & VeriKVC & 1.32 & 99.71 & 96.14 & 95.67 \\
3 & Keystroke Wizards & 1.78 & 99.59 & 95.9 & 94.04 \\
4 & U-CRISPER & 2.44 & 99.38 & 95.54 & 91.43 \\
5 & YYama & 2.54 & 99.27 & 95.38 & 89.61 \\
6 & BiDA Lab & 2.71 & 99.26 & 95.31 & 89.81 \\
7 & Challenger$\ast$ & 2.8 & 99.22 & 95.23 & 89.64 \\
8 & BioSense & 5.17 & 98.23 & 93.51 & 81.16 \\
\bottomrule
\end{tabular}
\vspace{0.2cm}

\begin{tabular}{c|c|c|c|c|c|c|c}
\multicolumn{8}{c}{\textbf{Mobile}} \\
\bottomrule
\textbf{Position} & 
\textbf{Team} & 
\makecell{\textbf{EER (\%)}$\downarrow$} &
\makecell{\textbf{FNMR @0.1\%}\\ \textbf{FMR (\%)$\downarrow$}} & 
\makecell{\textbf{FNMR @1\%}\\ \textbf{FMR (\%)$\downarrow$}} & 
\makecell{\textbf{FNMR @10\%}\\ \textbf{FMR (\%)$\downarrow$}} &
\textbf{AUC (\%)$\uparrow$} &
\textbf{Accuracy (\%)$\uparrow$} \\
\bottomrule
1 & LSIA & 3.61 & 63.62 & 17.44 & 0.6 & 99.28 & 96.39 \\
2 & VeriKVC & 3.78 & 65.88 & 18.39 & 0.95 & 99.04 & 96.22 \\
3 & Keystroke Wizards & 5.83 & 84.14 & 41.58 & 1.93 & 98.34 & 94.17 \\
4 & U-CRISPER & 8.76 & 94.79 & 67.15 & 6.68 & 96.54 & 91.24 \\
5 & YYama & 4.16 & 69.62 & 24.41 & 0.72 & 99.09 & 95.84 \\
6 & BiDA Lab & 9.45 & 94.77 & 67.67 & 8.53 & 96.22 & 90.55 \\
7 & Challenger$\ast$ & 5.19 & 77.46 & 32.89 & 1.55 & 98.69 & 94.81 \\
8 & BioSense & 11.83 & 88.11 & 60.48 & 14.43 & 94.83 & 88.17 \\
\bottomrule
\end{tabular}
\vspace{0.2cm}

\begin{tabular}{c|c|c|c|c|c}

\multicolumn{6}{c}{Mean Per-Subject Distributions} \\
\bottomrule
\textbf{Position} & \textbf{Team} &
\textbf{EER (\%)$\downarrow$} & \makecell{\textbf{AUC}\\  \textbf{(\%)$\uparrow$}} & \makecell{\textbf{Accuracy}\\  \textbf{(\%)$\uparrow$}} & \makecell{\textbf{Rank-1}\\\textbf{(\%)$\uparrow$}} \\
\bottomrule
1 & LSIA & 1.03 & 99.76 & 96.24 & 96.11 \\
2 & VeriKVC & 1.35 & 99.64 & 96.09 & 94.64 \\
3 & YYama & 1.66 & 99.53 & 95.88 & 92.5 \\
4 & Challenger$\ast$ & 2.17 & 99.39 & 95.58 & 91.11 \\
5 & Keystroke Wizards & 2.66 & 99.15 & 95.23 & 88.05 \\
6 & U-CRISPER & 5.07 & 97.86 & 93.37 & 73.94 \\
7 & BiDA Lab & 5.25 & 97.89 & 93.28 & 75.92 \\
8 & BioSense & 5.75 & 97.78 & 92.95 & 78.53 \\
\bottomrule
\multicolumn{6}{l}{Please see footnote 14.}
\end{tabular}
\label{tab:all_metrics}
\end{table*}

%% file: 6_Experimental_Results.tex
\subsection{Biometric Verification}
\label{subsec:biometric_verification}
The results of the experiments are reported in Table \ref{tab:all_metrics}. The table is divided into two parts, each one corresponding to one scenario: desktop and mobile. Each half can be further divided into the two cases considered: results obtained in the global genuine and impostor distributions (see Sec. \ref{sec:experimental_protocol}), and results considering the mean values obtained for per-subject genuine and impostor distributions. Each row shows a different system, while the different metrics are reported along the columns. In all tables, the arrow next to the metric indicates whether it is desirable to obtain higher (arrow up) or lower (arrow down) values.

As can be seen in the table, five teams achieve better results in the desktop task in comparison with the mobile task (LSIA, Keystroke Wizards, U-CRISPER, BiDA Lab, BioSense), while the remaining three (VeriKVC, YYama, Challenger) perform better in the mobile task. Consequently, the KVC-onGoing results reflect the typical lower variability associated with the desktop task due to a more constrained acquisition scenario (i.e., in contrast to mobile devices, subjects are more likely to be sitting down and in a still position while typing on a desktop keyboard). Furthermore, the best result is achieved in the desktop task with a global EER of 3.33\%.

The best performing team of both tasks, LSIA, achieves a global EER of 3.33\% and 3.61\% respectively in the desktop and mobile task, with a solid gap with respect to the second best team of both tasks, VeriKVC, especially in the desktop task (3.33\% vs. 4.03\%). It is interesting to point out that both teams use more sophisticated loss functions in comparison with other teams. In fact, the LSIA team adopts an extension of the SetMargin loss \cite{2022_PR_SetMargin_Morales} with an additive penalty term combined with a learning curriculum of increasing difficulty, whereas the VeriKVC team applies the Additive Angular Margin Loss (ArcFace) \cite{deng2019arcface}. Such approaches outperform current state-of-the-art performance on the same KVC-onGoing experimental conditions, evidencing the importance of optimal loss functions for training \cite{stragapede2023kvc}. Moreover, the performance gap between these two teams might be given by the complexity of the model architecture adopted, i.e., the LSIA team uses a dual-branch (recurrent and convolutional) embedding model for distance metric learning with attention modules, whereas the VeriKVC uses an attention module as feature extractor followed by a CNN. More in-depth analyses combining the two architectures could be carried out in the future to shed light on which are the most impactful aspects in learning discriminative features.

\begin{figure*}[!t]
\begin{subfigure}{.5\textwidth}
  \centering
  \includegraphics[width=\linewidth]{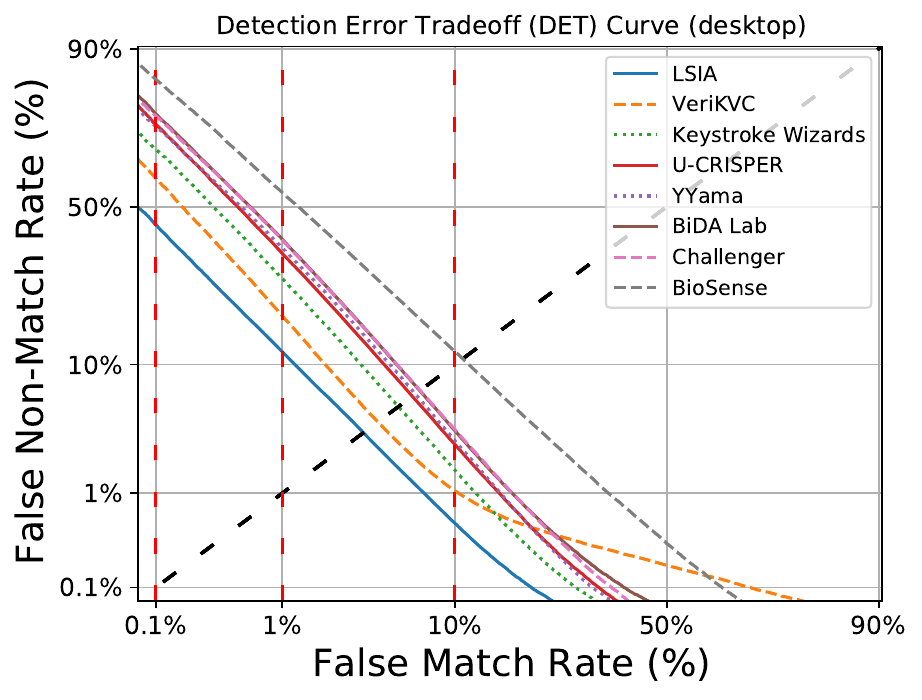}
  \caption{Desktop task.}
  \label{fig:det_desktop}
\end{subfigure}%
\begin{subfigure}{.5\textwidth}
  \centering
  \includegraphics[width=\linewidth]{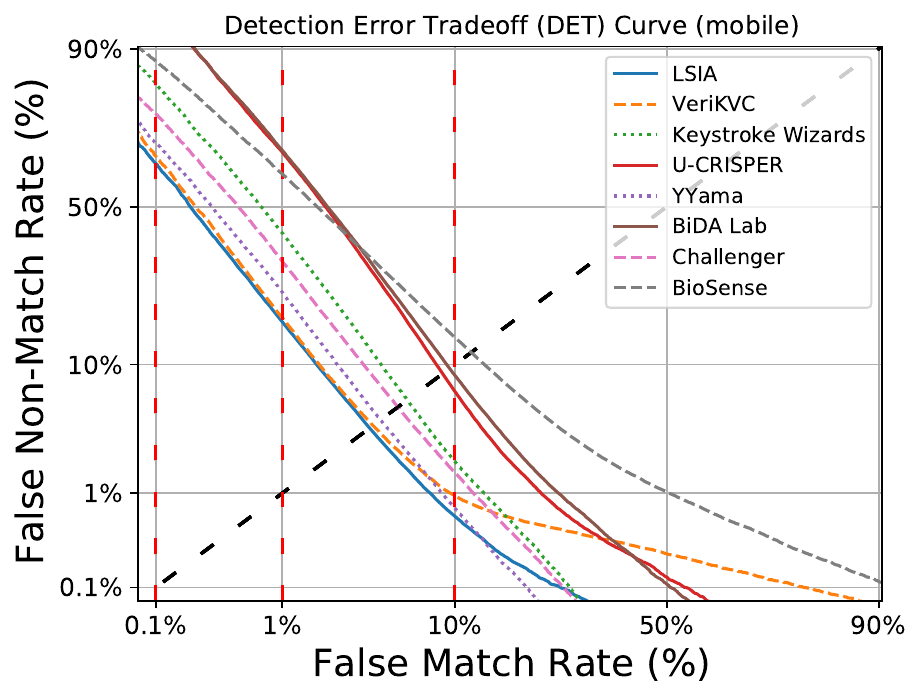}
  \caption{Mobile task.}
  \label{fig:det_mobile}
\end{subfigure}
\caption{\small DET curves including the results of all the biometric verification systems proposed in the KVC-onGoing challenge for both desktop and mobile tasks. The red dashed lines indicate the operational points 0.1\% FMR, 1\% FMR and 10\% FMR whereas the black dashed line indicate the points where the FMR = FNMR, corresponding to the EER.}
\label{fig:dets}
\end{figure*}

Another interesting aspect to point out is related to the fact that the ranking in both desktop and mobile tasks is consistent across the two evaluation scenarios considered, i.e., global EER and mean per-subject EER. The adoption of per-subject EER is generally associated with better results as different thresholds are considered per subject, allowing the system to better adapt to each subject. That is, by tuning the acceptance threshold on the subject's enrolment sessions, the performance could increase significantly. In fact, in terms of the mean per-subject EER, the system developed by the LSIA team achieves results under 1\% in the desktop task (0.77\%), and very close to that in the mobile task (1.03\%). In general, using a dedicated threshold (mean per-subject EER) instead of a global one (global EER), the error rates drop by 50\% to 80\% roughly. Moreover, we observe that the best performing systems also achieve a greater error reduction.

The ranking is also consistent across the different metrics. For example, examining different threshold operating points (FNMR @0.1\% FMR, FNMR @1\% FMR, and FNMR @10\% FMR), as well as the AUC and verification accuracy, it is possible to see that the separation between the performance of different systems is clear. 

Concerning the other teams, the performance achieved is very satisfactory in both desktop and mobile tasks as well, with global EER results between 5.22\% (Keystroke Wizards) and 10.85\% (BioSense) for the desktop task and between 4.16\% (YYama) to 11.83\% (BioSense) for the mobile task. Considering the variety of the architectures proposed (RNNs, CNNs, Transformers), it is possible to conclude that biometric verification based on KD is a problem that can be tackled from several directions with good results. In particular, approaches based on distance metric learning (triplet and contrastive loss functions) used by Keystroke Wizards, U-CRISPER, YYama, BiDA Lab, and Challenger, seem to work better than cross-entropy loss (BioSense).

Comparing the two tasks, it is possible to see that, apart
from the first two positions, the rankings are different. In this analysis, it must be specified that the Keystroke Wizards team and BiDA Lab team are the only ones adopting a different approach in the two tasks, using a RNN in the desktop task, and a Transformer in the mobile task, leading to slightly different global EER results (5.22\% and 6.76\% in the desktop task vs. 5.83\% and 9.45\% for mobile). Regarding the YYama team, the proposed “dual-network” approach combines an embedding model for feature extraction and a classifier model for verification, achieving an absolute reduction of 2\% in global EER in the mobile task. In contrast, the opposite trend can be seen for the U-CRISPER team that proposes a GRU-Based Siamese Network, leading to 6.19\% global EER in the desktop task and 8.76\% in the mobile task. For this particular team, we point out that the performance highly degrades in the mobile task when fixing the threshold of the system to 1\% FMR, achieving a 67.15\% FNMR in comparison with the value of 60.78\% obtained by the BioSense team (which, in turn, has a worse global EER, 11.83\%). In general, the results obtained by teams from the third to the seventh position seem to suggest that approaches based on RNNs work better in the desktop task, while Transformer-based systems are better models on mobile devices. Such trend is in line with previous benchmarks \cite{stragapede2023kvc}.

For completeness, we include in Fig. \ref{fig:dets} a comparison of the biometric verification systems in terms of the Detection Error Tradeoff (DET) curves, for both desktop (Fig. \ref{fig:det_desktop}) and mobile (Fig. \ref{fig:det_mobile}) tasks. In the graphs, four operational points are marked: a black dashed line represents represents the points where the FMR = FNMR (i.e., the EER metric). In contrast, the three red lines represent two operational points in which the errors at the intersection with the DET curves are unbalanced. Specifically, they respectively represent the score of FMR @0.1\% FNMR, FMR @1\% FNMR, and FMR @10\% FNMR (also reported in Table \ref{tab:all_metrics}). 



Considering the DET curves, it is possible to observe that in the mobile task, three systems exhibit a change in the slope as the FMR increases (VeriKVC, U-CRISPER, BioSense), while a similar trend is visible in the desktop task only for VeriKVC. Having fixed the FNMR (considering a horizontal line on the DET curve graph), worse security of the system corresponds to higher values of FMR. Consequently, such trends highlight the deterioration of the biometric performance of a given system. Such trend is especially accentuated in the case of VeriKVC (based on a convolutional architecture), despite of being the second best system in terms of EER, as for a FNMR = 0.1\% the FMR is higher than 50\% for both desktop and mobile tasks.

\subsection{Fairness}

Table \ref{tab:demographic_metrics} shows all results of the fairness assessment provided throughout the proposed framework. Along the rows, the table is divided into two parts: the upper part presents the results in the desktop scenario, while the lower part is focused on the mobile scenario. Concerning the different metrics, it is necessary to point out that all metrics except $\textrm{SIR}$ are calculated considering 12 demographic groups, due to 6 age groups and 2 genders (see Sec. \ref{sec:resources}). In contrast, $\textrm{SIR}_{a}$ is computed considering scores divided by age only (square matrix of dimension 6), while $\textrm{SIR}_{g}$ considering gender only (square matrix of dimension 2), to keep the assessments of each of the two attributes independent (Sec. \ref{sec:metrics}).

\begin{table*}[!h]
\centering
\caption{\small Comparison of the presented keystroke biometric verification systems from the point of view of fairness metrics. STD: Standard Deviation of global accuracy across demographic groups; SER: Skewed Error Rate of global accuracy across demographic groups; FDR: Fairness Discrepancy Rate; IR: Inequity Rate; GARBE: Gini Aggregation Rate for Biometric Equitability; SIR: Skewed Impostor Rate with respect to age (SIR$_{a}$) and gender (SIR$_{a}$).}
\begin{tabular}{c|c|c|c|c|c|c|c}
\multicolumn{8}{c}{\textbf{Desktop}} \\
\bottomrule
\textbf{Experiment} & \makecell{\textbf{STD (\%)$\downarrow$}} & \makecell{\textbf{SER}$\downarrow$} & \textbf{FDR$\uparrow$} & \textbf{IR$\downarrow$} & \textbf{GARBE$\downarrow$}  & \textbf{SIR$_{a}$ (\%)$\downarrow$} & \textbf{SIR$_{g}$ (\%)$\downarrow$} \\
\bottomrule
LSIA & 0.641 & 1.025 & 97.061 & 2.079 & 0.132 & 4.032 & 3.045 \\
VeriKVC & 0.369 & 1.012& 97.019 & 1.553 & 0.08 & 2.482 & 1.834 \\
Keystroke Wizards & 0.541 & 1.022 & 96.921 & 1.45 & 0.065 & 2.858 & 2.29 \\
U-CRISPER & 0.574 & 1.022 & 95.801 & 1.455 & 0.064 & 2.474 & 1.976 \\
YYama & 0.637 & 1.026 & 94.485 & 1.486 & 0.074 & 21.031 & 16.302 \\
BiDA Lab & 0.602 & 1.023 & 95.234 & 1.39 & 0.056 & 2.919 & 2.229 \\
Challenger & 0.692 & 1.027 & 95.711 & 1.432 & 0.058 & 2.862 & 2.3 \\
BioSense & 0.924 & 1.035 & 95.265 & 1.383 & 0.06 & 3.146 & 2.405 \\
\bottomrule
Mean & 0.623 & 1.024 & 95.937 & 1.523 & 0.074 & 5.226 & 4.048 \\
\bottomrule
\vspace{0.2cm}
\end{tabular}
\begin{tabular}{c|c|c|c|c|c|c|c}
\multicolumn{8}{c}{\textbf{Mobile}} \\
\bottomrule
\textbf{Experiment} & \makecell{\textbf{STD (\%)$\downarrow$}} & \makecell{\textbf{SER}$\downarrow$} & \textbf{FDR$\uparrow$} & \textbf{IR$\downarrow$} & \textbf{GARBE$\downarrow$}  & \textbf{SIR$_{a}$ (\%)$\downarrow$} & \textbf{SIR$_{g}$ (\%)$\downarrow$} \\
\bottomrule
LSIA & 0.664 & 1.025 & 94.327 & 4.011 & 0.214 & 5.111 & 4.834 \\
VeriKVC & 0.67 & 1.023 & 95.73 & 2.288 & 0.169 & 4.488 & 4.276 \\
YYama & 0.662 & 1.023 & 94.897 & 2.592 & 0.148 & 43.435 & 43.812 \\
Challenger & 0.882 & 1.032 & 94.938 & 2.598 & 0.13 & 6.068 & 6.535 \\
Keystroke Wizards & 0.995 & 1.036 & 95.275 & 2.216 & 0.139 & 7.077 & 7.022 \\
U-CRISPER & 1.124 & 1.041 & 95.74 & 2.339 & 0.133 & 3.372 & 4.143 \\
BiDA Lab & 1.551 & 1.057 & 94.836 & 2.338 & 0.162 & 6.44 & 6.097 \\
BioSense & 1.362 & 1.051 & 89.252 & 3.138 & 0.168 & 4.061 & 3.962 \\
\bottomrule
Mean & 0.989 & 1.036 & 94.374 & 2.69 & 0.158 & 10.01 & 10.09 \\
\bottomrule
\end{tabular}
\vspace{0.2cm}
\label{tab:demographic_metrics}
\end{table*}

By comparing the two tasks, it is possible to observe that for all metrics, in most cases, the mobile scenario is more prone to show a less \textit{fair} performance. This is confirmed by the mean values of the results achieved by all systems, contained in the last row of the desktop and mobile table subsections. Consequently, more demographic information is retained and reflected in the scores for this task. In the literature, traditionally a higher variability is attributed to mobile touchscreens in comparison with desktop keyboards due to differences in the pose or activity of typing subjects. While the networks adopted show to be able to model well such variability achieving similar biometric recognition performance in the two cases, it appears that the differences due to demographic aspects are more pronounced in the mobile case.

The global STD considers the differences in the verification accuracy computed for each specific demographic group with the global EER threshold. The global SER represents the ratio between the highest and lowest accuracy from all demographic groups and it does not show the same correlation. 
Table \ref{tab:glob_eer_dem} breaks down how global STD and global SER are computed for the LSIA team. For brevity, only the results of one team (LSIA), are shown. The table contains the accuracy values obtained for each demographic group. An additional column reports the mean values across genders, while an additional row reports the mean values across age groups.
The STD and SER values reported on the bottom of each table subsections (STD: 0.641\%, SER 1.025\% for desktop; STD: 0.664\%, SER: 1.025\% for mobile) correspond to those in Table \ref{tab:demographic_metrics}. In both tasks, the accuracy values are lower for the the female category. Moreover, in terms of age, in the mobile case a clear gradient increasing along the rows is visible for males, while the desktop task does not exhibit a similar trend. 
Considering all teams, VeriKVC with a CNN-based architecture combined with ArcFace loss function are able to achieve the best performance according to the global SER (both tasks) and global STD (desktop).


FDR, IR and GARBE are computed based on genuine scores and impostor scores obtained from comparisons between different subjects belonging to the same demographic group (similar impostors only). The threshold for the global 1\%FMR is applied to the separate groups of scores obtaining different FNMRs and FMRs for each demographic group. Two parameters $\alpha$ and $\beta$ = 1 - $\alpha$ are used to express the level of concern applied to differences in FMR and FNMR respectively. By setting $\alpha$ = 0.5, the same importance is applied in this case to FMR and FNMR. In particular, while the FDR shows a trend that is in line with what is achieved by the previous metrics in the desktop scenario (LSIA and VeriKVC with the best results, and fairness correlated with the biometric verification ranking), a different behavior can be observed in the mobile task and for IR and GARBE in both tasks. Mobile FDRs all fall in a similar range, apart from the BioSense team (89.252). In contrast, the LSIA team has the worst IR in both tasks. This might be due to the fact that IR is based on ratios, and very small differences can lead to high ratio values if computed with error rates that are very low in absolute terms. A similar scenario occurs in the case of GARBE, which is in turn based on the Gini coefficients: in both tasks, the LSIA and VeriKVC teams show the worst performance in terms of fairness, whereas the distribution of the results of the remaining teams is more consistent.

Analyzing the results of the systems in terms of fairness so far, depending on the metric different trends can be observed. Specifically, the STD, SER, and FDR seems to be correlated with the ranking of the verification performance. Consequently, conclusions about fairness based on STD and SER should preferably be drawn between systems achieving similar biometric performance. Anyway, in light of the results achieved we can conclude that the differences between the different demographic groups considered are not very significant, for example comparing them with the case of other biometrics such as the face 
, in which soft-biometric attributes such as the ethnicity, age or gender of the subject are often immediately recognizable from the raw data. From this perspective, in light of the biometric verification performance achievable, KD can be considered a \textit{privacy-friendly} biometric characteristic.
\begin{table}[t]
\centering
\caption{\small LSIA team results in terms of global accuracy for the final evaluation datasets (desktop and mobile), evaluated for the different demographic groups (gender and age). STD refers to the standard deviation, whereas SER refers to the Skewed Error Rate (Sec. \ref{sec:metrics}). Both metrics are computed from all the elements of each table.}
\label{table:bias}

\begin{tabular}{c|ccc}
\multicolumn{4}{c}{\textbf{LSIA, Desktop, Global Accuracy (\%)}} \\
\bottomrule
 & \textbf{Male} & \textbf{Female} & \textbf{Mean} \\   
\hline
\textbf{10 - 13} & 96.69 & 95.94 & 96.32 \\
\textbf{14 - 17} & 96.64 & 96.26 & 96.45 \\
\textbf{18 - 26} & 96.77 & 96.52 & 96.64 \\
\textbf{27 - 35} & 96.59 & 96.22 & 96.4 \\
\textbf{36 - 44} & 96.03 & 95.7 & 95.86 \\
\textbf{45 - 79} & 96.17 & 94.42 & 95.3 \\
\textbf{Mean} & 96.44 & 95.82 & - \\
\bottomrule
\multicolumn{4}{c}{\textbf{STD}: 0.641\%, \textbf{SER}: 1.025}\\
\end{tabular}
\vspace{0.2cm}
\\
\begin{tabular}{c|ccc}
\multicolumn{4}{c}{\textbf{LSIA, Mobile, Global Accuracy (\%)}} \\
\bottomrule
 & \textbf{Male} & \textbf{Female} & \textbf{Mean} \\   
\hline
\textbf{10 - 13}  & 95.89 & 96.01 & 95.95 \\
\textbf{14 - 17} & 96.04 & 95.55 & 95.8 \\
\textbf{18 - 26}  & 96.43 & 94.99 & 95.71 \\
\textbf{27 - 35} & 96.49 & 96.06 & 96.28 \\
\textbf{36 - 44} & 96.94 & 96.22 & 96.58 \\
\textbf{45 - 79}  & 97.4 & 95.33 & 96.36 \\
\textbf{Mean} & 96.66 & 95.63 & - \\
\bottomrule
\multicolumn{4}{c}{\textbf{STD}: 0.664\%, \textbf{SER}: 1.025}\\
\end{tabular}

\label{tab:glob_eer_dem}
\end{table}

\begin{figure*}[t!]
\centering
  \includegraphics[width=0.6\linewidth]{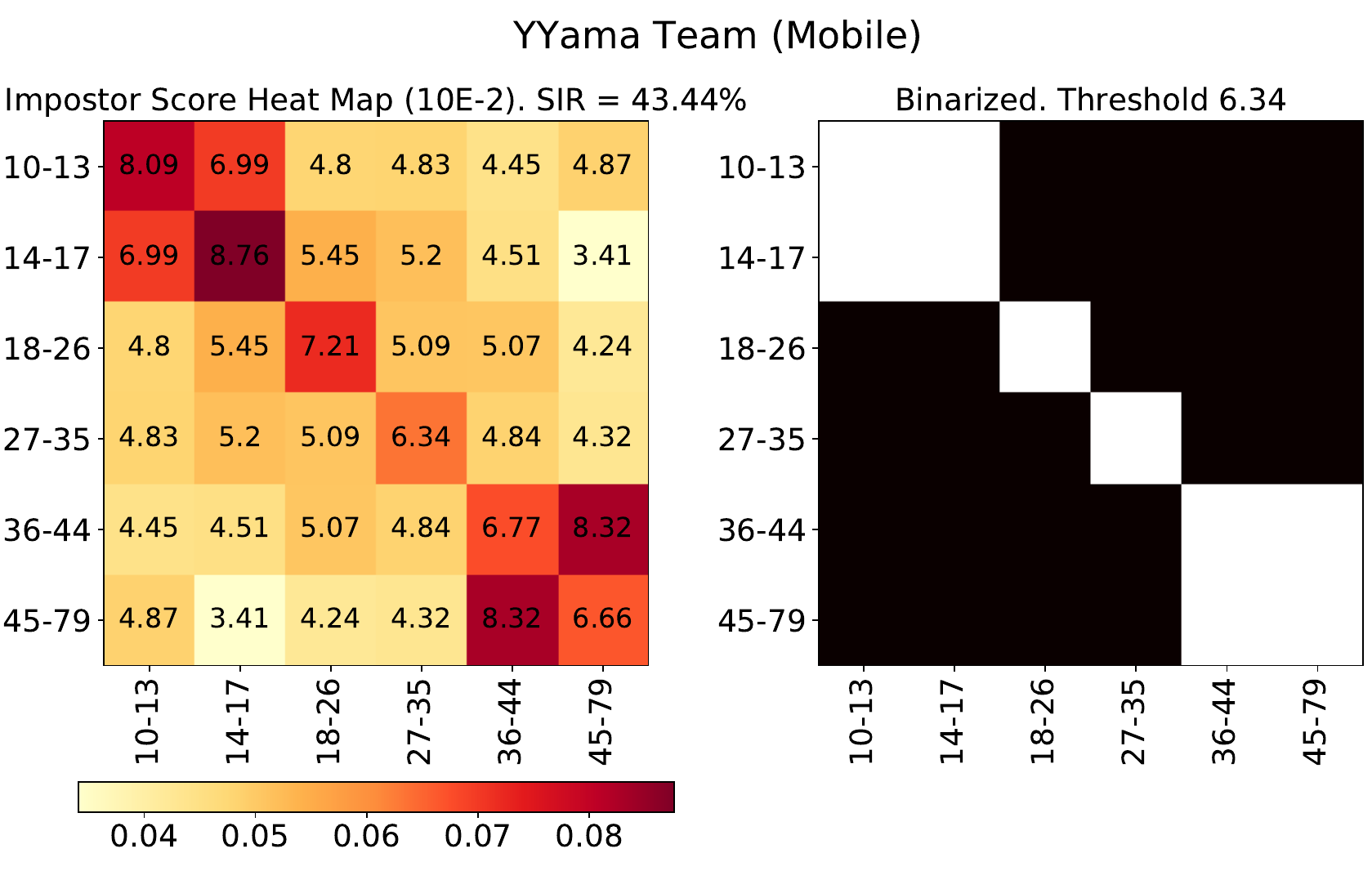}
  \vspace{0.5cm}\includegraphics[width=.6\linewidth]{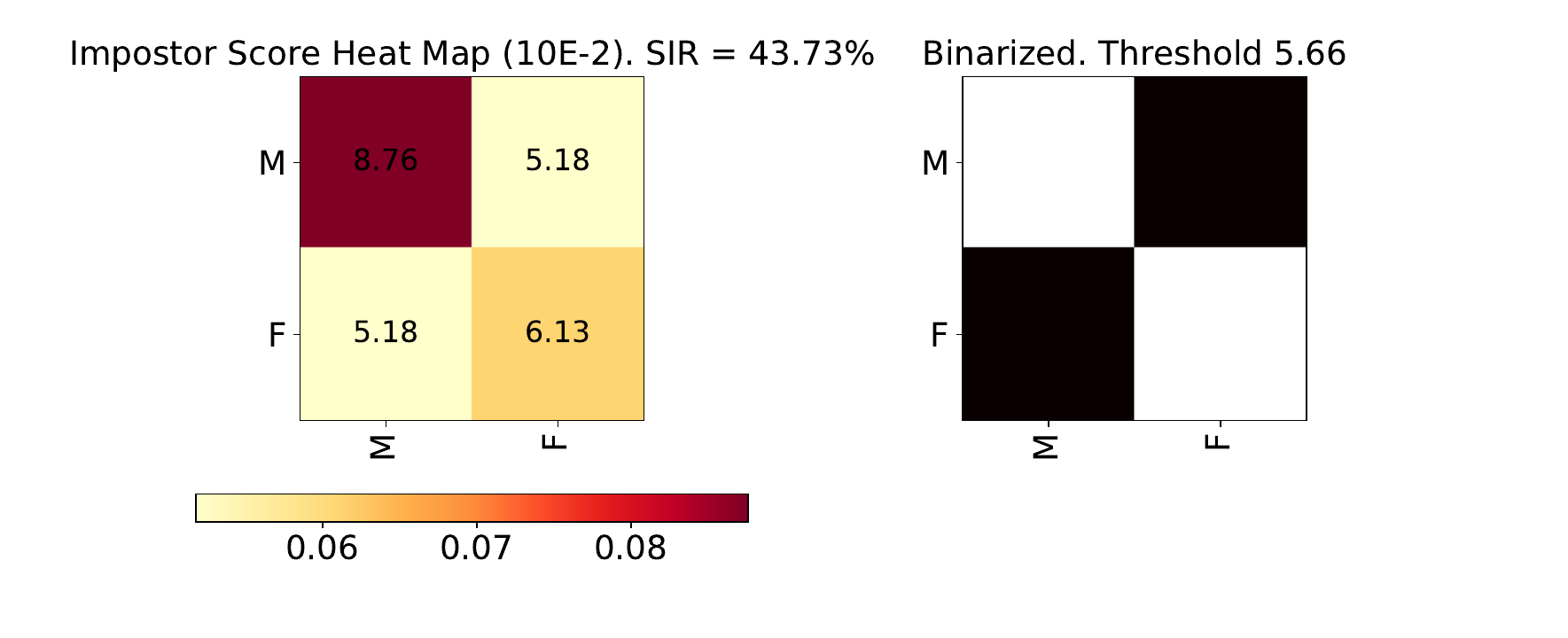}
\caption{\small Heat maps (normal and binarized) of the impostor scores of the YYama system with the highest $SIR_{a}$ (top) and $SIR_{g}$ (bottom) (YYama).}
\label{fig:heat_map}
\end{figure*}

The last two columns of the table show the SIR, respectively in terms of age and gender. This metrics are useful to quantify the demographic distribution skew of the scores between impostor comparisons, independently from the system threshold, and consequently how harder is, on average, a comparison depending on the demographic attributes of the compared subjects. Fig. \ref{fig:heat_map} illustrates its rationale in the form of heat maps for the case of the YYama team scores on the mobile task, for which the highest values of SIR are reported. The top half of the image contains the age heat maps, while the bottom half the gender heat maps. On the left, the heat maps are displayed on a scale of colors, while on the right the binarized version based on an automatic threshold is shown. For both tasks, it is possible to separate quite well the distribution of the scores of impostors belonging to the same demographic group (along the main diagonal) from those of impostors belonging to different demographic groups. This peculiar result might be due to the normalization technique or to the particular ``dual-network'' architecture employed.



%% file: 6a_Limitations.tex
We acknowledge some limitations of the generalizabilty of the results presented. As reported in Table \ref{tab:demographic_metrics} (Sec. \ref{sec:resources}), the datasets provided are not equally distributed with respect to age (all datasets) and gender (development sets only). This is due to the original subject distribution in the Aalto Keystroke Datasets, as balancing these categories would have entailed discarding a significant amount of subjects. To alleviate this aspect, we provide age and gender information of the subjects in the development sets, aiming to permit further considerations on the effects of demographic factors. Moreover, the majority of the subjects reported to be from the US, and English native speakers. This is not representative of the general population and might bias the data towards representing subjects interested in an online typing test (\textit{self-selection} bias). 

In addition, several other attributes were acquired through the final questionnaire proposed to the participants, such as the hands or fingers used, their typing experience, but also concerning the keyboard layout, orientation and possible auto-correction tools (for mobile devices), etc. However, we observed that in most cases, several values were missing, perhaps because the subjects did not complete the questionnaire. Moreover, other aspects that might influence the results might be the mood of the subjects while typing, their physical position, the short time interval between acquisition sessions, and the effect of a user changing the acquisition device. While these factors might all have an impact on the recognition performance, within the scope of this study we are deliberately targeting \textit{large-scale} analysis, making considerations about age and gender only. In light of the high number of factors to consider, we opted to maximize the amount of subjects and comparisons to validate the significance of the experimental results. More research is needed to disentagle these factors and to investigate their interactions considering specific subgroups.

%% file: 7_Conclusions.tex
This article has described the experimental framework and results of the Keystroke Verification Challenge - onGoing (KVC-onGoing)\footnote{\href{https://sites.google.com/view/bida-kvc/}{https://sites.google.com/view/bida-kvc/}}, a novel platform for the research community  to push forward state-of-the-art performance. Hosted on CodaLab\footnote{\href{https://codalab.lisn.upsaclay.fr/competitions/14063/}{https://codalab.lisn.upsaclay.fr/competitions/14063/}}, the proposed framework not only allows a complete assessment of the verification performance, but it also returns several metrics related to biometric fairness based on the comparison scores. 
The evaluation results have proved the high discriminative power of KD, reaching state-of-the-art results as low as 3.33\% of EER and 11.96\% of FNMR @1\% FMR in the desktop task, and 3.61\% of EER and 17.44\% of FNMR @1\% FMR in the mobile task. These results were achieved by the LSIA team by employing a dual-branch (recurrent and convolutional) embedding model with attention mechanism, trained with a custom loss function based on SetMargin loss \cite{2022_PR_SetMargin_Morales}. The second best team, VeriKVC, adopted a CNN-based model trained with ArcFace \cite{deng2019arcface}, achieving 4.03\% of EER and 18.79\% of FNMR @1\% FMR in the desktop task, and 3.78\% of EER and 18.39\% of FNMR @1\% FMR in the mobile task. In any case, most teams achieved less than 10\% of EER.
In terms of fairness, a baseline assessment of KD was performed for future studies. As a behavioral biometric characteristic, the analyzed scores showed relatively low levels of bias in comparison with physiological biometrics such as face. Nevertheless, the levels reported are not negligible: as an example, we report values for FDR, IR, GARBE respectively of 97.061, 2.079, 0.132 (desktop), and 94.327, 4.011, 0.214 (mobile) for the best team. Moreover, based on the SIR and impostor score heat map, a clear demographic correlation is revealed in the case of the YYama team, showing that specific patterns associated with age and gender in the raw data are reflected in the scores.

The next directions of research will go towards the optimization of the model architectures to improve the recognition performance \cite{type2branch}. With respect to past studies, to the best of our knowledge, for the first time a stricter operating point, FNMR @ 0.1\% FMR, is proposed for KD. By doing so, the differences between the biometric verification results are stretched and much clearer. 
In terms of demographic attribute assessment, further large-scale studies should focus on examining the differences in subjects' typing behavior due to biological, cultural, or linguistic factors, to further analyze the bias and patterns in the typing behavior of people. In this sense, new findings might be of great interest for several branches of the HCI community, e.g. privacy protection \cite{delgadosantos2021survey}, continuous user authentication \cite{stragapede2023behavepassdb}, security of minors online \cite{BORJ2023110039}, user experience improvement \cite{dunlop2022text2030}, etc.

%% file: 0_Main.bbl
\begin{thebibliography}{10}

\bibitem{delgado2023exploring}
Paula Delgado-Santos, Ruben Tolosana, Richard Guest, Farzin Deravi, and Ruben Vera-Rodriguez.
\newblock Exploring transformers for behavioural biometrics: A case study in gait recognition.
\newblock {\em Pattern Recognition}, 143:109798, 2023.

\bibitem{stragapede2022ijcb}
Giuseppe Stragapede, Ruben Vera-Rodriguez, Ruben Tolosana, and et~al.
\newblock {IJCB 2022} mobile behavioral biometrics competition {(MobileB2C)}.
\newblock In {\em Proc. IEEE Int. Joint Conf. on Biometrics (IJCB)}, 2022.

\bibitem{stragapede2023behavepassdb}
Giuseppe Stragapede, Ruben Vera-Rodriguez, Ruben Tolosana, and Aythami Morales.
\newblock {BehavePassDB: public database for mobile behavioral biometrics and benchmark evaluation}.
\newblock {\em Pattern Recognition}, 134:109089, 2023.

\bibitem{TOLOSANA2022108609}
Ruben Tolosana and Ruben Vera-Rodriguez \textit{et al.}
\newblock {SVC-onGoing: Signature verification competition}.
\newblock {\em Pattern Recognition}, 127, 2022.

\bibitem{mandryk2023combating}
Regan L et~al. Mandryk.
\newblock {Combating Toxicity, Harassment, and Abuse in Online Social Spaces: A Workshop at CHI 2023}.
\newblock In {\em Ext. Abstr. of the 2023 CHI Conf. on Human Factors in Computing Systems}, pages 1--7, 2023.

\bibitem{BORJ2023110039}
Parisa~Rezaee Borj, Kiran Raja, and Patrick Bours.
\newblock Online grooming detection: A comprehensive survey of child exploitation in chat logs.
\newblock {\em Knowledge-Based Systems}, 259:110039, 2023.

\bibitem{morales2020keystroke}
Aythami Morales, Alejandro Acien, Julian Fierrez, John~V Monaco, Ruben Tolosana, Ruben Vera, and Javier Ortega-Garcia.
\newblock Keystroke biometrics in response to fake news propagation in a global pandemic.
\newblock In {\em IEEE Annual Computers, Software, and Applications Conference (COMPSAC)}, pages 1604--1609, 2020.

\bibitem{haaretz}
Omer Benjakob.
\newblock {Netanyahu vs. Israeli Security Chiefs: Wikipedia Is New Front in Gaza War Blame Game}.
\newblock {\em Haaretz}, 2023-11-17.

\bibitem{2021_TTS_Biases_Terhorst}
Philipp Terhorst, Jan~N. Kolf, Marco Huber, Florian Kirchbuchner, Naser Damer, Aythami Morales, Julian Fierrez, and Arjan Kuijper.
\newblock A comprehensive study on face recognition biases beyond demographics.
\newblock {\em IEEE Trans. on Technology and Society}, 3(1):16--30, March 2022.

\bibitem{8966639}
Abeer A.~N. Buker, Giorgio Roffo, and Alessandro Vinciarelli.
\newblock {Type Like a Man! Inferring Gender from Keystroke Dynamics in Live-Chats}.
\newblock {\em IEEE Intelligent Systems}, 34(6):53--59, 2019.

\bibitem{10.1007/978-3-319-91189-2_33}
Avar Pentel.
\newblock Predicting user age by keystroke dynamics.
\newblock In Radek Silhavy, editor, {\em Artificial Intelligence and Algorithms in Intelligent Systems}, pages 336--343, 2019.

\bibitem{doi:10.1080/0144929X.2014.907343}
A.F.M. Nazmul~Haque Nahin, Jawad~Mohammad Alam, Hasan Mahmud, and Kamrul Hasan.
\newblock Identifying emotion by keystroke dynamics and text pattern analysis.
\newblock {\em Behaviour \& Information Technology}, 33(9):987--996, 2014.

\bibitem{telecom4030021}
Ioannis Tsimperidis, Denitsa Grunova, Soumen Roy, and Lefteris Moussiades.
\newblock Keystroke dynamics as a language profiling tool: Identifying mother tongue of unknown internet users.
\newblock {\em Telecom}, 4(3):369--377, 2023.

\bibitem{Dhakal2018}
Vivek Dhakal, Anna~Maria Feit, Per~Ola Kristensson, and Antti Oulasvirta.
\newblock Observations on typing from 136 million keystrokes.
\newblock In {\em Proc. CHI Conf. on Human Factors in Computing Systems}, 2018.

\bibitem{palin2019people}
Kseniia Palin, Anna~Maria Feit, Sunjun Kim, Per~Ola Kristensson, and Antti Oulasvirta.
\newblock How do people type on mobile devices? {Observations} from a study with 37,000 volunteers.
\newblock In {\em Proc. Int. Conf. on Human-Computer Interaction with Mobile Devices and Services}, 2019.

\bibitem{kboc}
Aythami Morales, Julian Fierrez, Ruben Tolosana, Javier Ortega-Garcia, Javier Galbally, Marta Gomez-Barrero, André Anjos, and Sébastien Marcel.
\newblock Keystroke biometrics ongoing competition.
\newblock {\em IEEE Access}, 4:7736--7746, 2016.

\bibitem{stragapede2023kvc}
Giuseppe Stragapede, Ruben Vera-Rodriguez, Ruben Tolosana, Aythami Morales, Naser Damer, Julian Fierrez, and Javier Ortega-Garcia.
\newblock {Keystroke Verification Challenge (KVC): Biometric and Fairness Benchmark Evaluation}.
\newblock {\em IEEE Access}, 2023.

\bibitem{bigdata}
{G. Stragapede, R. Vera-Rodriguez, R. Tolosana \textit{et al.}}
\newblock {IEEE BigData 2023 Keystroke Verification Challenge (KVC)}.
\newblock {\em Proc. IEEE Int. Conf. on Big Data (in press)}, 2023.

\bibitem{typenet}
Alejandro Acien, Aythami Morales, John~V. Monaco, Ruben Vera-Rodriguez, and Julian Fierrez.
\newblock {TypeNet: Deep Learning Keystroke Biometrics}.
\newblock {\em IEEE Transactions on Biometrics, Behavior, and Identity Science}, 4(1):57--70, 2022.

\bibitem{typeformer}
Giuseppe Stragapede, Paula Delgado-Santos, Ruben Tolosana, Ruben Vera-Rodriguez, Richard Guest, and Aythami Morales.
\newblock Typeformer: Transformers for mobile keystroke biometrics.
\newblock {\em Neural Computing and Applications}, pages 1--15, 2024.

\bibitem{greyckeytroke}
Romain Giot, Mohamad El-Abed, and Christophe Rosenberger.
\newblock Greyc keystroke: a benchmark for keystroke dynamics biometric systems.
\newblock In {\em Proc. Int. Conf. on Biometrics: Theory, Applications, and Systems}, 2009.

\bibitem{cmu}
Kevin~S Killourhy and Roy~A Maxion.
\newblock Comparing anomaly-detection algorithms for keystroke dynamics.
\newblock In {\em Proc. Int. Conf. on Dependable Systems \& Networks}, 2009.

\bibitem{biosecurid}
Julian Fierrez and et~al.
\newblock {BiosecurID: a multimodal biometric database}.
\newblock {\em Pattern Analysis and Applications}, 13:235--246, 2010.

\bibitem{rhu}
Mohamad El-Abed, Mostafa Dafer, and Ramzi El~Khayat.
\newblock {RHU Keystroke: A mobile-based benchmark for keystroke dynamics systems}.
\newblock In {\em Proc. Int. Carnahan Conf. on Security Technology}, pages 1--4, 2014.

\bibitem{clarksonI}
Esra Vural, Jiaju Huang, Daqing Hou, and Stephanie Schuckers.
\newblock Shared research dataset to support development of keystroke authentication.
\newblock In {\em Proc. Int. Joint Conf. on Biometrics}, 2014.

\bibitem{sun2016shared}
Yan Sun, Hayreddin Ceker, and Shambhu Upadhyaya.
\newblock Shared keystroke dataset for continuous authentication.
\newblock In {\em Proc. IEEE Int. Workshop on Information Forensics and Security}, 2016.

\bibitem{clarksonII}
Christopher Murphy, Jiaju Huang, Daqing Hou, and Stephanie Schuckers.
\newblock Shared dataset on natural human-computer interaction to support continuous authentication research.
\newblock In {\em Proc. Int. Joint Conf. on Biometrics}, 2017.

\bibitem{acien2021becaptcha}
Alejandro Acien, Aythami Morales, Julian Fierrez, Ruben Vera-Rodriguez, and Oscar Delgado-Mohatar.
\newblock {BeCAPTCHA: Behavioral bot detection using touchscreen and mobile sensors benchmarked on HuMIdb}.
\newblock {\em Engineering Applications of Artificial Intelligence}, 98:104058, 2021.

\bibitem{ceker2017}
Hayreddin Çeker and Shambhu Upadhyaya.
\newblock {Sensitivity Analysis in Keystroke Dynamics using Convolutional Neural Networks}.
\newblock In {\em {Proc. Workshop on Information Forensics and Security}}, 2017.

\bibitem{Li2022}
Jianwei Li, Han-Chih Chang, and Mark Stamp.
\newblock {Free-Text Keystroke Dynamics for User Authentication}.
\newblock {\em Artificial Intelligence for Cybersecurity}, pages 357--380, 2022.

\bibitem{morales2022setmargin}
Aythami Morales, Julian Fierrez, Alejandro Acien, Ruben Tolosana, and Ignacio Serna.
\newblock {SetMargin loss applied to deep keystroke biometrics with circle packing interpretation}.
\newblock {\em Pattern Recognition}, 122:108283, 2022.

\bibitem{vaswani2017attention}
Ashish Vaswani and et~al.
\newblock Attention is all you need.
\newblock {\em Advances in neural information processing systems}, 30, 2017.

\bibitem{neacsu2023doublestrokenet}
Teodor Neacsu, Teodor Poncu, Stefan Ruseti, and Mihai Dascalu.
\newblock {DoubleStrokeNet: Bigram-Level Keystroke Authentication}.
\newblock {\em Electronics}, 12(20):4309, 2023.

\bibitem{reinecke2015labinthewild}
Katharina Reinecke and Krzysztof~Z Gajos.
\newblock Labinthewild: Conducting large-scale online experiments with uncompensated samples.
\newblock In {\em Proc. of the ACM Conf. on Computer Supported Cooperative Work \& Social Computing}, 2015.

\bibitem{wobbrock2007measures}
Jacob~O Wobbrock.
\newblock Measures of text entry performance.
\newblock {\em Text entry systems: Mobility, accessibility, universality}, pages 47--74, 2007.

\bibitem{9507539}
Tiago de~Freitas~Pereira and Sébastien Marcel.
\newblock Fairness in biometrics: A figure of merit to assess biometric verification systems.
\newblock {\em IEEE Transactions on Biometrics, Behavior, and Identity Science}, 4(1):19--29, 2022.

\bibitem{grother2019face}
Patrick Grother, Mei Ngan, and Kayee Hanaoka.
\newblock {\em Face recognition vendor test (fvrt): Part 3, demographic effects}.
\newblock National Institute of Standards and Technology Gaithersburg, MD, 2019.

\bibitem{howard2022evaluating}
John~J Howard, Eli~J Laird, Yevgeniy~B Sirotin, Rebecca~E Rubin, Jerry~L Tipton, and Arun~R Vemury.
\newblock Evaluating proposed fairness models for face recognition algorithms.
\newblock {\em arXiv:2203.05051}, 2022.

\bibitem{gonzalez2021shape}
Nahuel Gonz{\'a}lez and et~al.
\newblock On the shape of timings distributions in free-text keystroke dynamics profiles.
\newblock {\em Heliyon}, 7(11), 2021.

\bibitem{gonzalez2022towards}
{N. Gonz{\'a}lez, E. P. Calot, J. S. Ierache, W. and Hasperu{\'e}}.
\newblock Towards liveness detection in keystroke dynamics: Revealing synthetic forgeries.
\newblock {\em Systems and Soft Computing}, 4:200037, 2022.

\bibitem{2022_PR_SetMargin_Morales}
Aythami Morales, Julian Fierrez, Alejandro Acien, Ruben Tolosana, and Ignacio Serna.
\newblock {SetMargin Loss applied to Deep Keystroke Biometrics with Circle Packing Interpretation}.
\newblock {\em Pattern Recognition}, 122:108283, February 2022.

\bibitem{deng2019arcface}
Jiankang Deng, Jia Guo, Niannan Xue, and Stefanos Zafeiriou.
\newblock {Arcface: Additive angular margin loss for deep face recognition}.
\newblock In {\em Proc. of the Conf. on computer vision and pattern recognition}, pages 4690--4699, 2019.

\bibitem{senerath2023behaveformer}
Dilshan Senerath and et~al.
\newblock {BehaveFormer: A Framework with Spatio-Temporal Dual Attention Transformers for IMU enhanced Keystroke Dynamics}.
\newblock In {\em Int. Joint Conf. on Biometrics}, 2023.

\bibitem{watanabe2023tree}
Shuhei Watanabe.
\newblock Tree-structured parzen estimator: Understanding its algorithm components and their roles for better empirical performance.
\newblock {\em arXiv:2304.11127}, 2023.

\bibitem{yang2014multimodal}
Qing Yang and et~al.
\newblock A multimodal data set for evaluating continuous authentication performance in smartphones.
\newblock In {\em Proc. of the ACM Conf. on Embedded Network Sensor Systems}, pages 358--359, 2014.

\bibitem{gonzalez}
Nahuel Gonzalez, Enrique~P. Calot, and Jorge~S. Ierache.
\newblock A replication of two free text keystroke dynamics experiments under harsher conditions.
\newblock In {\em Int. Conf. of the Biometrics Special Interest Group (BIOSIG)}, pages 1--6, 2016.

\bibitem{type2branch}
Nahuel Gonz{\'a}lez, Giuseppe Stragapede, Rub{\'e}n Vera-Rodriguez, and Rub{\'e}n Tolosana.
\newblock {Type2Branch: Keystroke Biometrics based on a Dual-branch Architecture with Attention Mechanisms and Set2set Loss}.
\newblock {\em arXiv:2405.01088}, 2024.

\bibitem{delgadosantos2021survey}
Paula Delgado-Santos, Giuseppe Stragapede, Ruben Tolosana, Richard Guest, Farzin Deravi, and Ruben Vera-Rodriguez.
\newblock {A Survey of Privacy Vulnerabilities of Mobile Device Sensors}.
\newblock {\em ACM Comput. Surv.}, 2022.

\bibitem{dunlop2022text2030}
Mark~D Dunlop and et~al.
\newblock {TEXT2030-Shaping Text Entry Research in 2030}.
\newblock In {\em Adj. Proc. of the Int. Conf. on Human-Computer Interaction with Mobile Devices and Services}, 2022.

\end{thebibliography}
